\documentclass[a4paper]{exam}

\usepackage[utf8]{inputenc} % padrão UTF8
\usepackage[T1]{fontenc} % hifenização correta no Português
\usepackage{amsthm} % \theoremstyle
\usepackage{amsmath} % \dfrac, \numberwithin
\usepackage{amssymb} % \Finv, \mathbb{R}
\usepackage{enumerate} % enumerate
\usepackage{float} % \begin{table}[H]
\usepackage{graphicx} % gráficos
\usepackage{subfig} % substitui o finado 'subfigure', mescla com R
\usepackage{multirow} % Para centralizar os elementos das colunas
\usepackage{multicol}
\usepackage{hyperref} % Sumário clicável and CC
\hypersetup{colorlinks=false,pdfborderstyle={/S/U/W 1},pdfborder=0 0 1} % permite underline nos links \_
\usepackage{Zallman} % Tipo de letra Capital
\usepackage{lettrine} % Primeira letra Capital

\LettrineTextFont{\itshape}
\usepackage{tcolorbox} % para caixas coloridas
\usepackage{tikz} % draw rectangle
\usetikzlibrary{positioning} % new position of objects
\usetikzlibrary{patterns} % preamble
\tcbuselibrary{skins} % preamble
\usepackage{ulem} % \sout, strike text (techado)
\usepackage{wasysym} % \smiley, notas musicais, masculino e feminino
\usepackage{wrapfig} % floating figures
\usepackage{authblk} % para indicar afiliação dos autores
\usepackage[super]{nth} % \nth{1}, \nth{2}, \nth{3}, \nth{4}
\usepackage[yyyymmdd]{datetime} % date format
\usepackage{booktabs} % beautiful tables
\newcommand{\ra}[1]{\renewcommand{\arraystretch}{#1}}

\usepackage{xcolor}
\usepackage{pagecolor}
\pagecolor{black}
\color{white}
\hypersetup{
    colorlinks,
    citecolor=white,
    filecolor=white,
    linkcolor=white,
    urlcolor=gray,
    anchorcolor=white
}

\usepackage[pscoord]{eso-pic}

\makeatletter
\patchcmd{\ttlh@hang}{\parindent\z@}{\parindent\z@\leavevmode}{}{}
\patchcmd{\ttlh@hang}{\noindent}{}{}{}
\makeatother

\graphicspath{{./img/}}

\theoremstyle{plain}
\newtheorem{theorem}{Theorem}
\newtheorem{example}{Example}

\usepackage{hyphenat}
\begin{document}

\title{\huge{D E C A D E S} \\ \medskip
       \footnotesize{O F} \\
       \huge{J U R I M E T R I C S}}
\author[1]{Filipe J. Zabala\thanks{filipe.zabala@pucrs.br}}
\author[2]{Fabiano F. Silveira\thanks{fabiano.feijo.silveira@gmail.com}}  
\affil[1]{\href{http://www.pucrs.br/technology/}{School of Technology}, \href{http://www.pucrs.br/en}{PUCRS}}

\renewcommand\Authands{ and }
  \date{2019-12-31 \\
    \begingroup
    \fontsize{9pt}{11pt}\selectfont
    \endgroup}
  \maketitle

%------------------------------------------------------------------------------------------------------------------------------%
\begin{flushright}
\texttt{
  Jurimetria é harmonia \\ 
  entre o exato e o direito \\ 
  calibra o que é meu com os fatos \\ 
  previsível, enfim, quem diria?
}
\end{flushright}
%--------------------------------------------------------------------------------------------------------------------------------------------------%

\tableofcontents

\newpage
\begin{abstract}

  Jurimetrics: decades of history, decades to-be auspicious. A Brazilian point of view on the trajectory of this forgotten concept in the quantitative approach of the law, with code and examples in free software.

\end{abstract}

\section{Old Wine in New Bottles}
\lettrine[lines=2, slope=0.6em]{T}{he year} 2019 is interesting in a `jurimetrical' point of view. Several historical decades have been completed, which makes an appropriate moment for reflections on possible deviations that may be distorting the current understanding of the matter. 

The conceptual framework of jurimetrics was first presented $(2019-1949)/10 = 7$ decades ago by \cite{loevinger1949jurimetrics}\footnote{ \url{https://mncourts.libguides.com/lee_loevinger}}. His manifesto begins considering that `one of the greatest anomalies of modern times that the law, which exists as a public guide to conduct, has become such a recondite mystery that it is incomprehensible to the public and scarcely intelligible to its own votaries'. Even nowadays the scenario is not much different, but the tools available at this time allow to address this issue, rather than simply  `make the people obey the laws they do not understand'. 

In Loevinger's writings from 1949 to 1992 it is possible to observe the influence of a wide variety of philosophers, like Oliver Wendell Holmes Jr., Thomas Bayes, Francis Bacon, Aristotle, among others. He was `standing on the shoulders of giants', as the saying attributed $(2019-1675)/10 \approx 34$ decades ago to Isaac Newton. This idea, however, was documented $(2019-1159)/10 = 86$ decades ago in John of Salisbury's Metalogicon \cite{salisbury2009metalogicon}, which accredits to the French philosopher Bernard of Chartres. On the foreword of \cite{merton1993shoulders}, Humberto Eco suggests that this concept is found in the texts of Priscian in the \nth{5} century, around 160 decades ago.

\begin{quote}
\textit{As Merton himself says, Bernard's Aphorism is known through its quotation by John of Salisbury in the Metalogicon (I can confirm his answer to a question raised by Merton: it is, indeed, in III, 4), and Bernard is not the original inventor, for the concept (if not the metaphor of the dwarfs) appears in Priscian six centuries earlier.} (Humberto Eco on the foreword of \cite{merton1993shoulders}, p. xiv.)
\end{quote}

Even in modern science is not straightforward to measure the contribution of each collaborator on a complex theory, although there are proposals in the literature \cite{winston1985suggested}. \cite{donoho201550years} discusses \textit{$\left[t\right]$he 50 years of data science}, when more than 5 decades ago John Tukey wrote \textit{The Future of Data Analysis}, pointing an `unrecognized science whose subject of interest was learning from data, or `data analysis''. `No scientific discovery is named after its original discoverer', states the \textit{Stigler's Law of Eponymy} \cite{stigler1980stigler} -- which ironically is an eponym\footnote{ \url{https://en.wikipedia.org/wiki/Eponym}} -- in the sense that
\begin{quote}

\textit{Laplace employed Fourier transforms in print before Fourier published on the topic, that Lagrange presented Laplace transforms before Laplace began his scientific career, that Poisson published the Cauchy distribution in 1824, 29 years before Cauchy touched on it in an incidental manner, and that Bienaymé stated and proved the Chebychev inequality a decade before and in greater generality than Chebychev’s first work on the topic.} \cite[p. 148]{stigler1980stigler}
\end{quote}

In the text that assigned the term \textit{Googol}\footnote{ The term was coined by Edward Kasner's nephew, Milton Sirotta (1911–1981). According to \cite{koller2004origin} the name `Google' is a corruption of `Googol', misspelled by Sean Anderson in September of 1997 when the domain \href{https://google.com}{google.com} was registered.} to the number $10^{100}$, \cite{kasner1940mathematics} points out in the Chapter 1 -- \textit{New Names for Old} -- that

\begin{quote}
\textit{$\left[e\right]$very once in a while there is house cleaning in mathematics. Some old names are discarded, some dusted off and refurbished; new theories, new additions to the household are assigned a place and name.}  \cite[p. 3]{kasner1940mathematics}
\end{quote}

Thus, we are only guardians of the old wine, passed from bottle to bottle throughout history.

\subsection{Recognizing the bottles}
During research the authors were led to \cite{leibniz1666dissertatio}, that for the given purposes will be considered the milestone on the formal association of quantitative thought and law, although \cite{friendly2008guerry} refers to early 1660s texts concerning a `political arithmetic'. Leibniz sheds light on the subject focusing on symbolic representation of languages using basic components units. With more than thirty five decades old, his writings suggests `the primitive terms comprise things and, as well, modes or relations', in a translation from Latin \cite{amunategui2014symbolic}. This kind of structure involving small blocks associated by a set of rules is merged on the foundations of the legal reasoning and all kinds of quantitative approach like statistics, mathematics, computer science, data science, machine learning, artificial intelligence and many other labels.

\begin{figure}[!h]
  \begin{center}
    \includegraphics[width=0.8\textwidth]{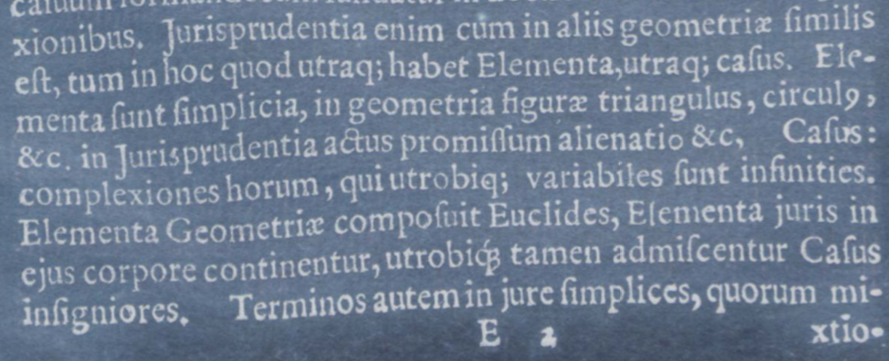}
  \caption{\cite{leibniz1666dissertatio} excerpt, p. 27.}
  \end{center}
\end{figure}

Exactly $(2019-1709)/10 = 31$ decades ago, Nicolaus (Nicolas/Nikolas/Niklaus) I Bernoulli (1687-1759) published his thesis \textit{De Usu Artis Canjectandi in Jure}, or `On the Use of the Art of Conjecturing in Law' \cite{bernoulli1709dissertatio}. It is a well-documented reference of the quantitative approach in the law, which deals with everyday issues such as lottery and insurance pricings, inheritance, confidence in witnesses, probabilities of innocence and survival of people. \cite{kohli1975kommentar} and \cite{hald2003history} points out he had great influence from the work of his uncle and master James (Jacques/Jakob/Jacob/Jacobi) I Bernoulli (1654-1705), \textit{Art of Conjecture}\footnote{ Part IV, Chapter II.}, published posthumously in 1713 \cite{bernoulli1713ars}.
\begin{quote}
\textit{An edition of the works of Jakob Bernoulli would be incomplete if one did not attach to it the thesis of his nephew Niklaus from the year 1709 `On the Use of the Art of Conjecturing in Law'. The spiritual father of this work is clearly Jacob. Whole sections from both the diary and the `Ars Conjectandi' Niklaus has taken literally. At other points, questions and mere hints of Jacob were taken up and further processed.} \cite[p. 541]{kohli1975kommentar}\footnote{ ``\textit{Eine Ausgabe der Werke von Jakob Bernoulli wäre unvollständig, würde man ihr nicht die Dissertation seines Neffen Niklaus aus dem Jahre 1709 `Über den Gebrauch der Mutmaßungskunst in Fragen des Rechts' beifügen. Der geistige Vater dieses Werkes ist eindeutig Jakob. Ganze Abschnitte sowohl aus dem Tagebuch als auch aus der `Ars Conjectandi' hat Niklaus wörtlich übernommen. An andern Stellen wurden Fragestellungen und bloße Andeutungen Jakobs aufgegriffen und weiterverarbeitet.}''}
\end{quote}

Also inspired on James I Bernoulli's work, \cite{condorcet1785essay} presents the currently known \textit{Condorcet's jury theorem}, a milestone in problems concerning voting systems. It can be applied in a wide variety of fields, such as social sciences and machine learning. The theorem can be stated in terms of a dichotomous variable assuming the values \texttt{1} and \texttt{0}, and it is considered reasonable to assign a `correct' or `incorrect' classification. If a decision maker -- e.g. a judge or classifier -- assigns correctly the \texttt{1}'s with probability $\theta$ greater than 1/2, the theorem asserts that more decision makers increases the overall probability of correct assignments. With $\theta$ less than 1/2, more decision makers decreases the overall probability of correct assignments, and for $\theta = 1/2$ the number of decisiors is indifferent. Adapted from \cite{berg1996condorcet}, the theorem is formally described as follows.

\newpage
\begin{theorem} (\textbf{Condorcet's theorem}) Let ($X_1, \ldots, X_n$) be $n$ independent binary distributed random variables such that $Pr(X_i = 1) = \theta > 1/2$ and $P_n = Pr(\sum X_i > n/2)$. Then (a) $P_n > \theta$ and (b) $P_n$ is monotonously increasing in $n$ and $P_n \longrightarrow 1$ as $n \longrightarrow \infty$. If $\theta < 1/2$, then $P_n < \theta$ and $P_n \longrightarrow 0$ as $n \longrightarrow \infty$. Finally, when $\theta = 1/2$, then $P_n = 1/2$ for all $n$.
\end{theorem}

\begin{example} If $n = 3$ and $\theta=0.6$ then $P_{3}$ -- the probability of at least two of three decision makers agree in the correctly assignment of \texttt{1}'s -- is given by \[P_{3} = 0.6 \times 0.6 \times 0.6 + 0.6 \times 0.6 \times 0.4 + 0.6 \times 0.4 \times 0.6 + 0.4 \times 0.6 \times 0.6 = 0.648 > 0.6.\]
\end{example}
\smiley

\begin{example} If $n = 3$ and $\theta=0.3$, then \[P_{3} = 0.3^3 + 3 \times 0.3^2 \times 0.7 = 0.216 < 0.3.\]
\end{example}
\smiley

\begin{example} If $n = 3$ and $\theta=0.5$, then \[P_{3} = 0.5^3 + 3 \times 0.5^3 = 0.5.\]
\end{example}
\smiley

\cite{condorcet1785essay} also points James I Bernoulli and Abraham de Moivre as precursors of the idea of seeking the probability of future events according to the law of past events, even they have given no method in their works to achieve this\footnote{ ``\textit{L'idée de chercher la probabilité des évènements futurs d'après la loi des évènements passés, parroît s'être présentée à Jacques Bernoulli \& à Moivre, mais ils n'ont donné dans leurs ouvrages aucune méthode pour y parvenir. M.$^{rs}$ Bayes \& Price en ont donné une dans les Transactions philosophiques, années 1764 \& 1765, \& M. de la Place est le premier qui ait traité cette question d'une maniére analytique.''}}. Considering the method, Condorcet points out the work of Thomas Bayes and Richard Price \cite{bayes1763essay}, as well as Pierre-Simon Laplace for treating the question analytically \cite{laplace1774memoire}. The Condorcet's theorem and its consequences has been discussed and extended in literature\footnote{ \cite{grofman1983thirteen}, \cite{boland1989majority}, \cite{ladha1992condorcet}, \cite{berg1993condorcet}, \cite{berg1996condorcet}, \cite{austen1996information}, \cite{list2001epistemic}, \cite{williams2004condorcet}, \cite{gehrlein2006condorcet}, \cite{gehrlein2011voting}, \cite{kaniovski2011optimal}, \cite{zaigraev2012note} and \cite{gottlieb2015voting}.} until recently.

Still in France, the \nth{19} century brought the works of Adolphe Quetelet, André-Michel Guerry and Simeón Poisson. They were interested in Laplace's work, as well in the analysis of conviction rates. During the development of \textit{l'homme moyen}\footnote{ Average man.} concept, from 1827 to 1835 Quetelet looked for potentially meaningful relationships in social data. According to \cite{stigler1986history},

\begin{quote}
\textit{$\left[h\right]$e examined birth and death rates by month and city, by temperature, and by time of day. He calculated the month of conception from the birth month and tried to relate it to marriage statistics. He investigated mortality by age, by profession, by locality, by season, in prisons, and in hospitals. He considered other human attributes: height, weight, growth rate, and strength. Quetelet's interests also extended to moral qualities: statistics on drunkenness, insanity, suicides, and crime.} \cite[p. 186]{stigler1986history}
\end{quote}

Based on \textit{Compte Général de L'administration de la Justice Criminelle en France} data \cite{france1829compte}, Quetelet worked in the analysis of conviction rates, also handled by \cite{guerry1833essai} and \cite{poisson1837recherches}. According to \cite{stigler1986history},

\begin{quote}
\textit{Quetelet had given the numbers of accused and convicted as 7,234 and 4,594; Poisson gave them as 6,652 and 4,037, respectively. Thus for Quetelet the 1825 conviction rate was 0.635, for Poisson it was 0.607. The explanation for this discrepancy is that in preparing the report for 1827 the minister of justice (Count Portalis) had discovered that in the report for 1825 the figures given had been augmented by those for accused condemned in absentia, and he had provided the needed correction. Apparently Quetelet missed this change (announced in a footnote on p. v of the Compte général for 1827), but Poisson did not.} \cite[p. 188]{stigler1986history}
\end{quote}

%%%% Plot
\begin{figure}[ht]
\begin{center}
\includegraphics[width=0.7\textwidth]{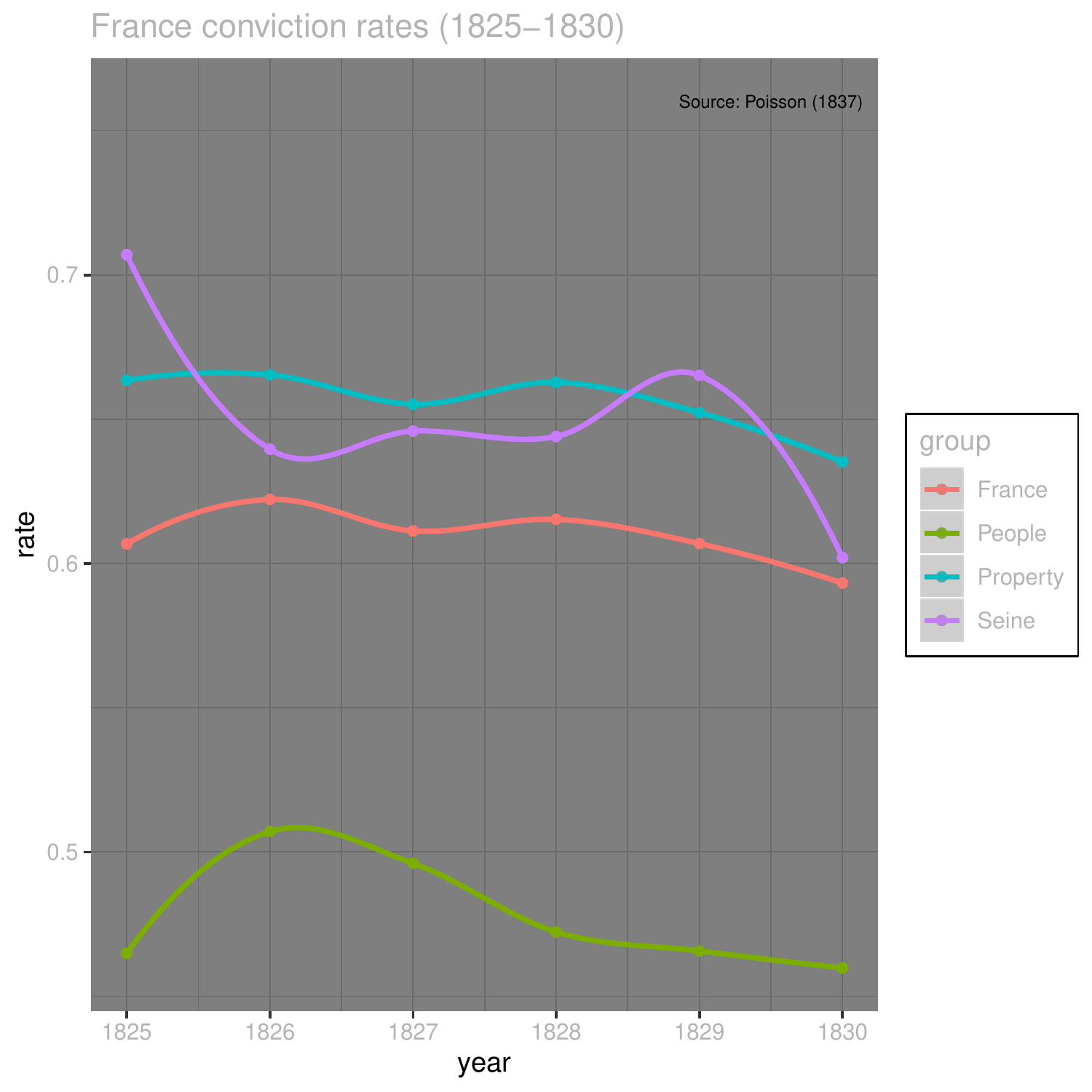}
\caption{France conviction rates (1825-1830)}
\label{fig:poisson}
\end{center}
\end{figure}

The seminal works of \cite{holmes1881common} and \cite{holmes1897path} lead to new perspectives in the Law, considering the precedent (the decided, or the \textit{stare decisis}) in similar cases. Discussing early forms of liability, Holmes declares `other tools are needed besides logic'.
\begin{quote}
\textit{The life of the law has not been logic: it has been experience.} \cite[p. 5]{holmes1881common}
\end{quote}
The \textit{Bayesian} approach provides a formal mechanism to incorporate information to decision maker's experience. Such approach -- discussed briefly in Sections \ref{sec:drinking}, \ref{sec:data} and \ref{sec:tps} -- also provides tools to prediction, suggested by Holmes Jr. in stating that
\begin{quote}
\textit{$\left[t\right]$he object of our study, then, is prediction, the prediction of the incidence of the public force through the instrumentality of the courts.} \cite[p. 1]{holmes1897path}
\end{quote}

\cite{llewellyn1930realistic} discusses the problem of defining law and defends the Roscoe Pound's `precepts', in opposition to `rules' for being a `term sufficiently ambiguous'. At page 460 Llewellyn considers `making the study of law a study in first instance of particularized situations', where `generalization must come from a resynthesis of such particularized studies'.

Inspired by Oliver Wendell Holmes Jr.'s writings and considering the advances in quantitative methods, \cite{loevinger1949jurimetrics} is a manifesto in defense of rationality in the law. Labeled \textit{jurimetrics}, this house cleaning occured seven decades ago and still provokes high-level discussions. Forward Llewellyn's next step, Loevinger identifies Roscoe Pound -- dean of Harvard Law School -- as `the most prominent spokesman' of Legal Realists, considered by Loevinger a `school of Sociological Jurisprudence' that `has developed more from the stimulus of (Rudolph von) Jhering than of Holmes'. Loevinger defends the `legal principle as the most significant aspect of law', opposed to the `emphasis on the particular case', `although agreeing with the Realists that law must adapt itself to social needs'.
\begin{quote}
\textit{`The scholars whose writings have emphasized the particular case, rather than the general rule, as the basis for a study of the law, have been called Legal Realists'.} \cite[p. 10]{loevinger1949jurimetrics}
\end{quote}

\begin{quote}
\textit{The next step forward in the long path of man's progress must be from jurisprudence (which is mere speculation about law) to jurimetrics -- which is the scientific investigation of legal problems.} \cite[p. 31]{loevinger1949jurimetrics}
\end{quote}

After the 1949 seminal article, Lee Loevinger was proliferous writing about science and law in the 1950s and 1960s. \cite{loevinger1950semantics} makes a book review of \textit{Courts on Trial: Myth and Reality in American Justice}, published by Jerome Frank in 1949. He points out the shift of Frank's ideas between 1930 and 1949, criticizing the `individualized cases' concept and the lack of a formal definition for `justice'. 

\cite{loevinger1952introduction} discusses some Holmes Jr.'s ideas, like `the life of the law has not been logic: it has been experience'. He considers this a slogan `at least superficially misleading', but points out that Holmes Jr. wish `a more conscious and rational recognition of the grounds of judicial decision'. Also makes a defense of the use of logic in law, referring indirectly to the concept of \textit{coherence} \cite{leonard1980roles}, \cite{loschi2002coherence}. Loevinger retakes Jerome Frank's work, citing Mortimer Adler, Walter Wheeler Cook, Dennis Lloyd, Julius Cohen, William James, John Stuart Mill, John Dewey among others, making considerations about the foundations of probability theory.
\begin{quote}
\textit{While there appears to be a real difference between a `fair preponderance of the evidence' and `clear and convincing proof', there is no obvious indication that there is a similar differentiation between the latter phrase and `proof beyond a reasonable doubt'. At least some of this confusion might be avoided if courts were to adopt a terminology of probability logic. The conventional mode of representation calls for the use of 0 to indicate that a proposition is untrue or impossible, the use of 1.0 to indicate that a proposition is true or certain, and the use of intervening values to indicate the relative probability (or frequency) of its being true.} \cite[p. 495]{loevinger1952introduction}
\end{quote}

Loevinger finishes \textit{An Introduction to Legal Logic} stating
\begin{quote}
\textit{$\left[t\right]$he future of the law, and perhaps of society itself, will depend upon the ability of the legal professionals to develop and utilize patterns and forms of thinking in the law adequate to deal with the complex problems of modern life.} \cite[p. 522]{loevinger1952introduction}
\end{quote} 

\cite{loevinger1958facts} refers again to `legal realists' and takes up the issue of `individualized cases', making a deep discussion in the distinction between `evidence', `facts'/`opinions', `subjective'/`objective' and `perceptions'/`conclusions' concepts. He discusses admissibility and coherence, suggesting the social policy `be accomplished by establishing an appropriate rule relating to the degree of proof required, rather than by artificial rules excluding classes of evidence'. He ends the article declaring 
\begin{quote}
\textit{in order to function effectively at any level, every mind must take account of the learning and thinking that has preceded and must approach contemporary problems in terms of contemporary concepts.} \cite[p. 175]{loevinger1958facts}
\end{quote}

\cite{loevinger1961jurimetrics} discusses the advances of science, considering that `science and law have been linked in man's speech and thinking for centuries'. He points out Karl Pearson's idea that `the classification of facts, the recognition of their sequence and relative significance is the function of science'. From Oliver Wendell Holmes Jr. writings, Loevinger remembers that `an ideal system of law should draw its postulates and its legislative justification from science' and `for the rational study of the law the black-letter, man may be the man of the present, but the man of the future is the man of statistics and the master of economics'. The scientific advances cited by Loevinger are related largely to electronics in the field of data retrieval. He defines in a very mature way the concepts of \textit{hardware} and \textit{software}:
\begin{quote}
\textit{`Hardware' means simply the mechanical devices -- the physical machines -- that science has produced. `Software' means the intellectual systems of designs and concepts that have been produced.} \cite[p. 258]{loevinger1961jurimetrics}
\end{quote}

He alerts `that legal prediction is an activity in which lawyers, and for that matter citizens in all occupations, are commonly engaged', and that the legal profession has no choice unless adapt to new technologies `to retain its position of intellectual leadership'. To apply in a practical way he points out
\begin{quote}
\textit{$\left[t\right]$he branch of mathematics that appears to be of the most immediate practical utility in the fields of law and the behavioral sciences is statistics. There is much in statistics that is of present practical application in day-to-day legal problems and it has good claim to be included in every law school curriculum.} \cite[p. 262]{loevinger1961jurimetrics}
\end{quote}

\cite{loevinger1962occam} discusses the principle of parsimony in scientific thinking, known as \textit{Occam's Razor} and summarized in the maxim \textit{`entities should not be multiplied beyond necessity'}\footnote{ \textit{`Essentia non sunt multiplicanda praeter necessitate', William of Occam, \nth{14} century}.}. \cite{loevinger1963jurimetrics} retakes some ideas of Oliver Holmes Jr. and brings jurimetrics as a new science, considering jurisprudence as `the philosophy of law'. Examining the methodology of legal inquiry, Leovinger considers the electronic data retrieval as an interesting and seminal area of work. He details some principles of text manipulation and points out an \textit{association factor} purposed by \cite{stiles1961association}. Such factor assigns a relevance to related terms, `where $A$ is the number of documents indexed by one term; $B$ is the number of documents indexed by a second term; $f$ is the number of documents indexed by the combination of both terms; and $N$ is the total number of documents in the collection. If $AB$ is greater than $fN$ the association is negative'.

\begin{figure}[!h]
  \begin{center}
    \includegraphics[width=0.6\textwidth]{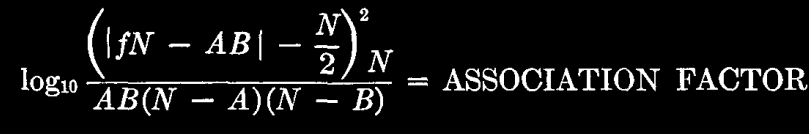}
  \caption{Association Factor by \cite{stiles1961association}.}
  \end{center}
\end{figure}

Loevinger also discusses a \textit{probabilistic indexing} suggested by \cite{maron1960relevance}, as well a system named `LEX' designed as a LEgal indeX by the Antitrust Division of the U.S. Department of Justice\footnote{ \url{https://www.justice.gov/atr}}. In the following excerpt he describes the modern approach to data manipulation:
\begin{quote}
\textit{Conceivably, it may become possible to record all the decisions in a field of law, or perhaps in the entire law, in such fashion that their full text can be searched electronically in microseconds.} \cite[p. 26]{loevinger1963jurimetrics}
\end{quote}

\begin{figure}[!h]
  \begin{center}
    \includegraphics[width=0.7\textwidth]{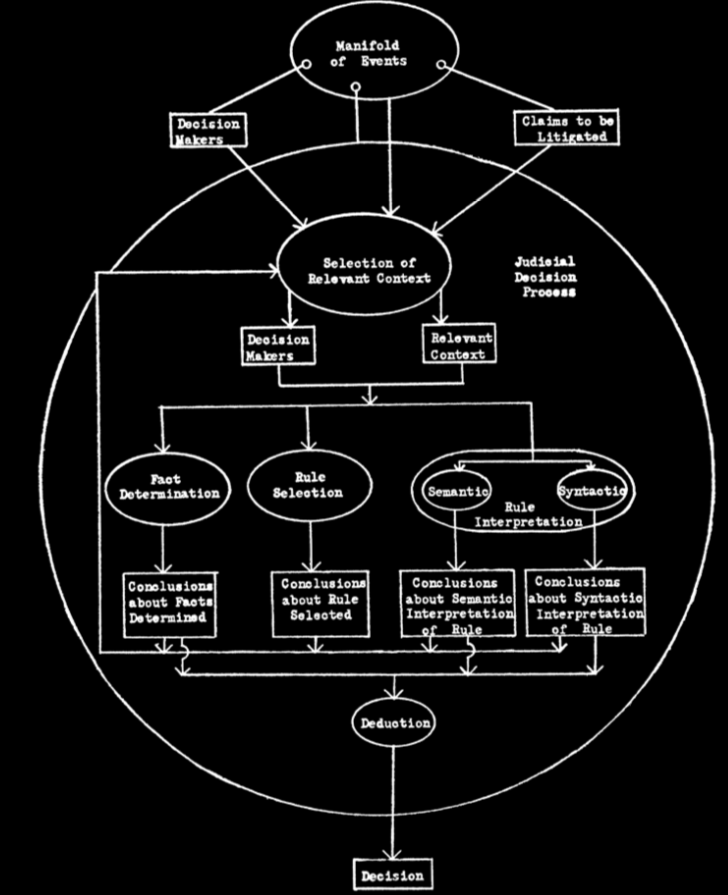}
  \caption{Scheme by \cite[p. 227]{allen1963modern}.}
  \label{fig:allen}
  \end{center}
\end{figure}

The discussion purposed by Loevinger is very opportune, considering that substantial textual structures are easily handled today with tools written in R, Python and REGEX (REGular EXpressions)\footnote{ \url{https://regex101.com/}}, for instance. The TF-IDF (Term Frequency--Inverse Document Frequency) \cite{sparck1972statistical} and its derived metrics \cite{lopes2016estimating} can be considered a new generation of relevance metrics discussed by Loevinger. Also driven by the modernization of the legal processes, \cite{allen1963modern}  considers schemes and diagrams as shown in Figure \ref{fig:allen}, while \cite{allen1963beyond} sets his thesis:
\begin{quote}
  \textit{If}
  \\ \smallskip
  $\;\;\;$ \textit{1. the written materials used in the tax field are more systematically drafted,}
  \\
  \textit{Then}
  \\ \smallskip
  $\;\;\;$ \textit{2. human beings will be able to `read' and `work with' those materials `better', and} 
  \\ \smallskip
  $\;\;\;$ \textit{3. automatic devices will be able to `read' and `work with' those materials `better'.}
  \begin{flushright} \cite[p. 714]{allen1963beyond}   \end{flushright}
\end{quote}

\cite{loevinger1966law} discusses science and law as rival systems, claiming at page 535 that `up to the present time law has made little use of science, or the empiric method, and (...) the lawyers generally have little understanding of science or its concepts and methods', pointing out a symposium issue in \textit{Law and Contemporary Problems}\footnote{\url{https://scholarship.law.duke.edu/lcp/vol28/iss1/}} as one of the few significant efforts to scientific work in the law. Loevinger also discusses the empirical approach given by \cite{kalven1966american} about jury behavior and makes a defense of the complementarity of methods and the difficulty in recognizing the limitations of each field.
\begin{quote}
\textit{The point that needs to be made, however, is that the empiric and the dialectic methods are not rivals or alternatives but complementary methods adapted to different problems and applicable in different situations. (...) Much of the difficulty in the relationship between law and science has arisen from the failure of both lawyers and behavioral scientists to recognize the limitations of their respective methods and the kinds of problems to which they are appropriate.} \cite[pp. 541-543]{loevinger1966law}
\end{quote}

Giving preference to data production over more theories, Loevinger considers `the records of judgements in the hundreds of trial courts' as `an obvious rich mine of data' and indicates `an ineluctable relationship between science and semantics'. He was concerned about the different understandings of vague terms as `justice', `reasonable' and `public interest', purposing `experimental and quantitative investigation of such semantic questions'. 

\begin{quote}
\textit{The profession of law and those it serves can hope and demand that the law schools begin to study both of the great systems of gathering data, the dialectic and the empiric, and bring both to bear in seeking solutions of the proliferating and increasingly complex problems of government in a scientific age, and in training those who will become our future governors.} \cite[p. 551]{loevinger1966law}
\end{quote}

\cite{loevinger1985science} makes a review of the main ideas of jurimetrics, in a lecture pointing out `the greatest developments of technology now appear to be tending toward a dispersion of power', in the sense that

\begin{quote}
\textit{the world today is divided between societies which value democracy and individual freedom and those which subjugate the individual to the state and are ruled by authoritarian regimes. Probably the most critical issue confronting us today is the distribution of power -- political power -- within our own society and in the community of nations. (...) In the critical area of communications new modes of transmitting and receiving information are proliferating and are segmenting audiences for all of the mass media. At the same time the new technologies provide increasing means for small groups and even individuals to disseminate views and make divergent voices heard.} \cite[p. 19]{loevinger1985science}
\end{quote}

\cite{loevinger1992standards} asserts both science and law `rest upon subjective judgments or assumptions', and `proof of facts in all disciplines rests upon subjective judgments of probability'. This is a Bayesian point of view, deeply supported in literature since \cite{bayes1763essay}.

\begin{figure}[!h]
  \begin{center}
    \includegraphics[width=0.7\textwidth]{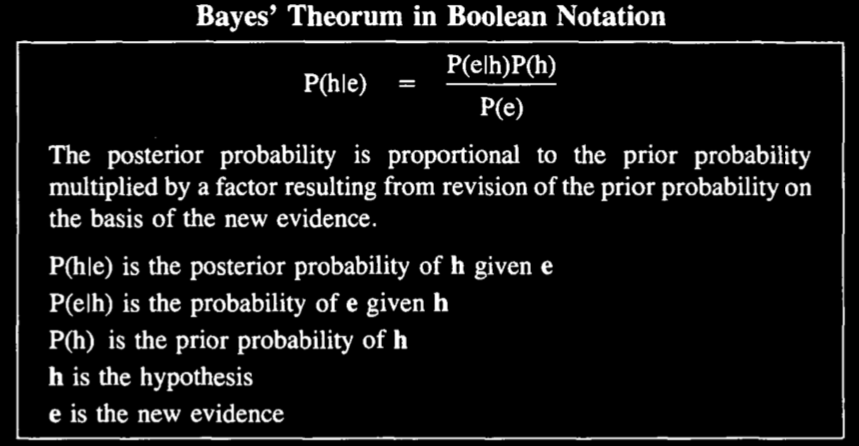}
  \caption{Bayes' Theorem by \cite[p. 327]{loevinger1992standards}}
  \label{fig:bayes}
  \end{center}
\end{figure}

An important reference concerning the subjectivist point of view of probability is the work of Bruno de Finetti. He purposed the \textit{representation theorem} \cite{definetti1937prevision}, where \textit{exchangeable variables}\footnote{An exchangeable variable is characterized when the order of the observations does not alter its probability law.} are \textit{conditionally independent} given a non-observed parameter $\theta$. This quantity is treated as a nuisance parameter, giving a predictive characterization to de Finetti's theorem. Throughout his work a strong subjectivist point of view becomes clear, as can be noticed, e.g., in the motto `PROBABILITY DOES NOT EXIST' \cite[p. x]{finetti1974theory} and in the following excerpt.
\begin{quote}
\textit{There is no way, however, in which the individual can avoid the burden of responsibility for his own evaluations. The key cannot be found that will unlock the enchanted garden wherein, among the fairy-rings and the shrubs of magic wands, beneath the trees laden with monads and noumena, blossom forth the flowers of `Probabilitas realis'. With these fabulous blooms safely in our button-holes we would be spared the necessity of forming opinions, and the heavy loads we bear upon our necks would be rendered superfluous once and for all.} \cite[p. 42]{finetti1975theory}
\end{quote}

Under this point of view, it is possible to formally assign the decision maker's opinion about $\theta$ to a probability distribution called \textit{prior}, which in turn is calibrated with data obtained from the \textit{likelihood function}. This procedure provides a \textit{posterior} distribution, or the decision maker's opinion about $\theta$ after the data. This approach can be naturally applied in the law, as pointed out by Loevinger.
\begin{quote}
\textit{Lawyers gather data, which they call `evidence'; scientists gather evidence, which they call `data'. Both terms mean the same thing, which is intellectual support for some conclusion or proposition.} \cite[p. 323]{loevinger1992standards}
\end{quote}

Loevinger retakes the Ockham's Razor concept from the Bayesian point of view \cite{jefferys1992ockham}, and points out legal rules of evidence and `burden of proof' conditioned on the admissibility of such proofs and evidences. He indicates some cases in which probability calculations was used by courts, discussing objective and subjective approaches to probability interpretation. Finally, \cite{loevinger1992logic} discusses philosophically some aspects of logic and Sociology, re-examinating `albeit somewhat reluctantly' \cite{kaye1992proof}, \cite{loevinger1992standards} and \cite{jasanoff1992judges}. 

\cite{kowalski1995legislation} points out the similarities between computing and the law, that `seem to cover all areas of computing software (...) as programming, program specification, database description and query, integrity constraints, and knowledge representation in artificial intelligence'. Besides that, he indicates the similarities between computing and law are `extend also to the problems that the two fields share of developing, maintaining and reusing large and complex bodies of linguistic texts'.
\begin{quote}
\textit{The linguistic style in which legislation is normally written has many similarities with the language of logic programming. (...) These extensions include the introduction of types, relative clauses, both ordinary negation and negation by failure, integrity constraints, metalevel reasoning and procedural notation.} \cite[p. 325]{kowalski1995legislation}
\end{quote}

At this point in the timeline it is possible to observe the non-intersection of Jurimetrics with Empirical Legal Research and Legal Realism. Several textbooks emerged indicating some point of views on the history of the quantitative empiricism in the law, to the best of our knowledge not quoting Loevinger and jurimetrics. \cite[$P=432, L=0, J=0$]{schlegel1995american}\footnote{$P$: number of (\#) pages. $L$: \# times \texttt{loevinger} term was found. $J$: \# times \texttt{jurimetric*} radical was found.} conceives \textit{American Legal Realism and Empirical Social Science} in the sense that `the point of this book is the stories'. The author declares to `love stories' and be `better at narrative than analytic history', offering to his readers `a capsule summary of \textit{the} story of Realism as it is usually told'.
\begin{quote}
\textit{Thus it was not until the 1920s that more than an isolated soul would claim that legal science was unscientific and so elicit clues as to why that was, and still is, the case. The group of scholars that made this claim and so brought the notion of science as an empirical inquiry if not into, then at least up against, law was the American Legal Realists.} \cite[p. 1]{schlegel1995american}
\end{quote}

\cite[$P=50, L=0, J=0$]{kritzer2009empirical} makes a bibliographic essay of the \textit{Empirical Legal Studies Before 1940},  stating that `\textit{Empirical Legal Studies} is a term that began to come into vogue around 2000', pointing out several studies conducted in the 1950s and early 1960s. He considers the legal realism as an essentially American movement `in significant part' because `almost all of the empirical legal research of the period was done by Americans focusing on the United States'. While the author do not discusses every study found, he claims to have `provided as complete a bibliography of that research as possible'. 

\cite[$P=1112, L=0, J=1$]{cane2010oxford}\footnote{The reference found was the citation of \textit{Robertson, D. (1982). ``Judicial Ideology in the House of Lords: A \textbf{Jurimetric} Analysis,'' British Journal of Political Science 12: 1–25.} Robertson does not cite Loevinger, considering the `celebrated paradigm is Schubert's \textit{The Judicial Mind}, a classic of what has been called `jurimetrics''.} indicates in \textit{The Oxford Handbook of Empirical Legal Research} that `in the American legal Academy, empirical research gained contemporary prominence in the late 1990s'. The authors points out `the genesis of empirically based studies of judicial behavior is commonly traced to the pioneering work of C. Hermann Pritchett in the late 1940s'.

In \textit{An Introduction to Empirical Legal Research}, \cite[$P=324, L=0, J=0$]{epstein2014introduction} retakes some Holmes Jr.'s ideas, as `the man of the future is the man of statistics and the master of economics', pointing out that the `future is here'. The authors present code and databases treated in R and Stata softwares. They use a classic statistical approach, considering methods that violates the likelihood principle\footnote{For more details see \cite{birnbaum1962foundations}, \cite{wechsler2008birnbaum} and \cite{mayo2014birnbaum}.} -- such as confidence intervals and p-value calculation from classical hypothesis tests -- as `close approximations to Bayesian methods', which can be noticed in the following excerpt.
\begin{quote}
\textit{Technically, to use population data to extrapolate to a different context or a future context, Bayesian statistics are the appropriate method. Bayesian statistics treat data as given and parameters as the things about which we have uncertainty. All the inferential tools described in this book are close approximations to Bayesian methods as long as we have sufficiently large samples, which provides justification for their use. In other words, it is appropriate to use hypothesis tests or confidence intervals as part of a process of extrapolation to different contexts or different time periods as an approximate Bayesian solution to the inferential problem.} \cite[p. 154]{epstein2014introduction}
\end{quote}

At last, the \textit{Empirical Legal Research} by \cite[$P=328, L=0, J=0$]{leeuw2016empirical} `covers all major fields of law, as the \textit{Oxford Handbook of Empirical Legal Research} (Cane and Kritzer, 2010) shows'. On the other hand, Loevinger exposes in his articles `considerable diversity among the references cited', discussing the ideas of a wide range of authors.
\begin{quote}
\textit{Some service has, indeed, been rendered by the modern thinkers, Bentham, Jhering, Holmes, Pound, the Realists, and others of similar views, in bringing law out of the sky and down to earth.} \cite[p. 15]{loevinger1949jurimetrics}
\end{quote}

\cite{bench2015knowledge} refers to \cite{loevinger1949jurimetrics} and \cite{mehl1958automation} as who anticipated many of the computational systems that have been implemented and proposed. For some reason, however, the Empirical Legal Research and Legal Realism literature does not gave the appropriate credit to Lee Loevinger. Clues about this issue are found (again) in \cite{loevinger1949jurimetrics}. At the end of the second paragraph of page 4, he asserts that `the trouble with law is not the public but the lawyers, that what is needed is not publicity but progress'. At pages 39-40 claims that `advertising is no substitute for research', and complements on the footnote \#85:
\begin{quote}
\textit{Needless to say, the term `research' is here used in its scientific sense, and what is carelessly called `legal research' by the average lawyer has no more relation to it than numerology has to statistics.} \cite[p. 40]{loevinger1949jurimetrics}
\end{quote}

Cultivating a collaborative thinking and inspired by known and unknown giants, the authors support a unified approach to the topic and celebrate the dedication of so many people for so many decades in building such a vast and detailed work. Cheers!

\subsection{Drinking the wine}\label{sec:drinking}

\begin{quote}
\textit{Of course it is not important what term is used to indicate the scientific discipline suggested. It is important that it have a distinctive name, as well as a general program. The name suggested here seems, to the author, as good as any, since it seems to indicate the nature of the subject matter, and corresponds to other similar terms, such as biometrics and econometrics.} \cite[p. 31]{loevinger1949jurimetrics}
\end{quote}

Jurimetrics is the application of quantitative methods in the law.\footnote{Albeit \cite[p. 8]{loevinger1963jurimetrics} asserts `(i)t is unnecessary, and perhaps impossible, to give a precise definition to the field of jurimetrics'.} So, anyone who considers quantitative approaches to legal problems is making jurimetrics. The term is concise and intuitive like biometrics, chemometrics, econometrics and so on. It combines the flexibility of the Human Science with the precision of the Natural Science. 

An example of this scientific association is the case \textit{United States v. Carroll Towing Co.}, concerning the sinking of the barge Anna C on January 4, 1944. Judge Learned Hand uses for the first time a cost-benefit analysis for assigning liability and determining negligence \cite{hand1947united}. Although \cite{feldman2005hand} discusses other approaches and indicates the called \textit{Hand rule} might produce games with inefficient equilibria, the case is a milestone in a modern use of quantitative methods in the law. In the judge's words, the Hand rule is described as follows.
\begin{quote}
\textit{Since there are occasions when every vessel will break from her moorings, and since, if she does, she becomes a menace to those about her; the owner's duty, as in other similar situations, to provide against resulting injuries is a function of three variables: (1) The probability that she will break away; (2) the gravity of the resulting injury, if she does; (3) the burden of adequate precautions. Possibly it serves to bring this notion into relief to state it in algebraic terms: if the probability be called $P$; the injury, $L$; and the burden, $B$; liability depends upon whether $B$ is less than $L$ multiplied by $P$: i.e., whether $B < LP$.} \cite[online]{hand1947united}
 \end{quote}

To the best of our knowledge, the first works in brazilian jurimetrics were produced in IME-USP\footnote{\textit{Instituto de Matemática e Estatística, Universidade de São Paulo} (Institute of Mathematics and Statistics, University of São Paulo).}. \cite{montoya1998prob} and \cite{montoya2001unconditional} presents a methodologic proposal for calculating the probability of paternity when DNA information available comes only from the accused's relatives. The proposed method allows to evaluate the probability of paternity considering the Hardy-Weinberg equilibrium law using the Bayesian approach. 

\cite{nakano2006novel} builds `a mathematical model for calculating the probability of paternity and implement it in software'.\footnote{\textit{Construir um modelo matemático para cálculo da probabilidade de paternidade e implementá-lo em software.} \cite[p. 5]{nakano2006novel}} The author also considers the Hardy-Weinberg equilibrium law, applying methods such Bayesian Networks and Full Bayesian Significance Test (FBST). Purposed by \cite{pereira1999evidence}, the FBST is a Bayesian measure of evidence for precise (sharp) hypotheses, as well as a Bayesian alternative to significance tests or, equivalently, to p-values.

\cite{wechsler2006analise} describes a technical report concerning a civil law issue of leasing contracts dollar-indexed for vehicle purchase. The authors studies aspects related to the validity -- from the vehicle buyers' financial point of view -- of a legal action against leasing companies. Making use of the Bayesian approach, they build a calculator based on a 2001 jurisprudency\footnote{Recurso Especial do STJ n$^{\circ}$. 473141.}, 27 variables and the opinion about the judgements. They presents the applications \textit{Check Balance}\footnote{Verifica Saldo.} and \textit{Decision Supporter}\footnote{Apoiador de Decisão.} in order to provide reference values to the decision maker. The opinion about how favorable are the legal decisions to the vehicle buyer or the leasing companies is given in a ordinal scale from 1 (highly favorable to vehicle buyer) to 5 (highly favorable to leasing companies), where 0 indicates ignorance about the decisions. To the levels of the ordinal scale are assigned parameters of a beta probability distribution \cite[pp. 210-275]{johnson1995continuous}.

\begin{figure}[!h]
  \begin{center}
    \includegraphics[width=0.6\textwidth]{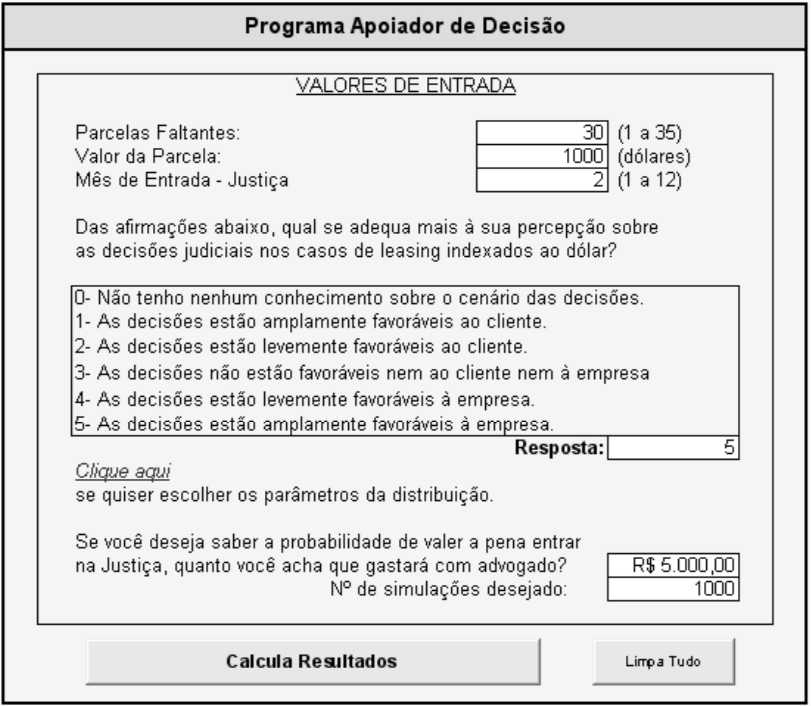}
  \caption{Decision Supporter window by \cite[p. 49]{wechsler2006analise}}
  \label{fig:bayes}
  \end{center}
\end{figure}

\cite{kadane2012probability} refers to a lecture given in the XI Brazilian Meeting on Bayesian Statistics, concerning the case of Kansas cellphone users\footnote{Quin Jackson et al. v. Sprint Nextel Corporation, Case No. 09-cv-2192 (N.D. Ill.).} in which `Sprint-Nextel was sued for conspiring with other cell phone providers to impose high prices for text-messaging'. Kadane discusses the probability sampling concept and its application, pointing out that in the present case `classical statistics did not address the court’s question, but Bayesian analysis did'.

\begin{figure}[!h]
  \begin{center}
    \includegraphics[width=7cm]{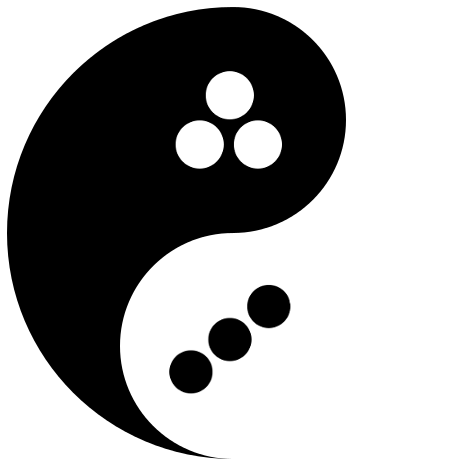}
  \caption{Juri-yang.}
  \label{jy}
  \end{center}
\end{figure}

Half decade ago \cite{zabala2014jurimetria} present a brief history of what was known so far about jurimetrics. Some considerations and suggestions on practical and theoretical applications of the subject are made. A division of jurimetrics into three prisms is proposed, formalizing a connection between law and quantitative thinking. Some theoretical and applied examples are presented, demonstrating part of the application in different contexts and raising fundamental questions for use in Brazilian law. In 26 April 2016 the same authors purposed a jurimetrics symbol\footnote{ \url{https://commons.wikimedia.org/wiki/File:Juri-yang.png}} called \textit{Juri-yang}. Presented in Figure \ref{jy}, it represents the harmony between `juri' -- indicated by three dots forming a stylized balance -- and `metrics', indicated by three aligned dots suggesting a stylized scatter plot. 

\cite{trecenti2015diagramas} presents the method  of influence diagram applied to civil justice litigation. The data were obtained from the web, through web scraping and text mining techniques. The R packages \texttt{tjsp} and \texttt{bnr} were developed, used respectively to facilitate access to the TJSP\footnote{\textit{Tribunal de Justiça de São Paulo.}} database and adjust the parameters of Bayesian networks based on a Gibbs sampler. The author discusses the importance of data extraction and manipulation, as well as the use of open-source packages that allow the reproducibility of jurimetric works.

\cite{nunes2016jurimetria} discusses how statistics can reinvent the law. The author cites a series of lectures taught in Brazil in 1973 by the philosophy professor at the Universities of Milan and Turin, Mario Losano. According the author, Losano `did not agree with the idea of the quantification of law and rejected statistics because he considered it to be incompatible with law in all its scope, including the study of general norms, principles and social values'\footnote{\textit{(...) não compactuava com a ideia da quantificação do Direito e rejeitava a Estatística por considerá-la incompatível com o Direito em toda sua abrangência, incluindo o estudo das normas gerais, princípios e valores sociais.} \cite[p. 104]{nunes2016jurimetria}}, considering that `predicting the content of a sentence would be impracticable'\footnote{\textit{(...) a previsão do conteúdo de uma sentença seria inviável.} \cite[p. 105]{nunes2016jurimetria}}. Considering the expression `jurimetrics', Nunes points out Lee Loevinger had a `deterministic view of knowledge'\footnote{\textit{(...) uma visão determinista do conhecimento.} \cite[p. 100]{nunes2016jurimetria}} and `stuck to the concept of deterministic causality, Loevinger understood that uncertainty would deprive the scientist of the means of identifying the causes of a phenomenon, preventing him from making predictions about his future behavior.'\footnote{\textit{Preso ao conceito de causalidade determinística, Loevinger entendia que a incerteza privaria o cientista dos meios de identificação das causas de um fenômeno, impedindo-o de formular previsões a respeito do seu comportamento futuro.} \cite[p. 100]{nunes2016jurimetria}}

\cite{marcondes2019assessing} discusses the necessity of `transparency and complete auditability' in judicial systems, considering random procedures to select jury, court or judge. The authors emphasizes `these principles are neglected by random procedures in some judicial systems, which are performed in secrecy and are not auditable by the involved parties'. As an example of such a procedure is discussed the assignment of cases in the Brazilian Supreme Court, presenting `a review of how sortition has been historically employed by societies and discusses how Mathematical Statistics may be applied to random procedures of the judicial system'. The authors finalize the article purposing a `statistical model for assessing randomness in case assignment', applied to the Brazilian Supreme Court.

\cite{aires2017norm}, \cite{aires2019automatic} and \cite{aires2019classification} develop an approach to identify potential conflicts between norms based on deontic logic. The authors evaluate the effectiveness of the presented techniques `using a manually annotated contract conflict corpus with results close to the current state-of-the-art for conflict identification, while introducing a more complex classification task of such conflicts' for which their method `surpasses the state-of-the art method'. \cite{zabala2019jurimetrics} present general purpose tools for jurimetrics, with code and databases  available in free software. Examples are given in Section \ref{sec:ex} and in the Figure \ref{fig:poisson}.

\subsection{\textit{Gueule de bois}}

% Contrary to their own analytical tradition, 
In 23 March 2019 was enacted in France the Law No. 2019-222, a regulation know as \textit{Article 33}. It provides a penalty up to 5 years of imprisonment, a fine up to 300,000 euros and possible loss of civil rights for anyone who discloses analysis of French judiciary data \cite{france2019article}\footnote{ ``\textit{Les données d’identité des magistrats et des membres du greffe ne peuvent faire l’objet d’une réutilisation ayant pour objet ou pour effet d’évaluer, d’analyser, de comparer ou de prédire leurs pratiques professionnelles réelles ou supposées. La violation de cette interdiction est punie des peines prévues aux articles 226-18,226-24 et 226-31 du code pénal, sans préjudice des mesures et sanctions prévues par la loi n$^{\circ}$. 78-17 du 6 janvier 1978 relative à l'informatique, aux fichiers et aux libertés.}''}. According to Article 33,

\begin{quote}
\textit{$\left[t\right]$he identity data of magistrates and members of court can not be reuse with the aim of object or effect of evaluating, to analyze, compare or predict their practices real or assumed. Violation of this prohibition is punished by the penalties provided for in Articles 226-18,226-24 and 226-31 of the Penal Code, without prejudice to the measures and sanctions provided for by Law n$^{\circ}$. 78-17 of 6 January 1978 relating to data processing, files and freedoms.}
\end{quote}

The penalties provided in Article 226-18 \cite{france2004article}\footnote{ ``\textit{Le fait de collecter des données à caractère personnel par un moyen frauduleux, déloyal ou illicite est puni de cinq ans d'emprisonnement et de 300 000 euros d'amende.''}} points out
\begin{quote}
\textit{$\left[t\right]$he collection of personal data by fraudulent, disloyal or unlawful means is punishable by five years' imprisonment and a fine of 300,000 euros.}
\end{quote}

The Article 226-31 \cite{france1994article}\footnote{ ``\textit{Les personnes physiques coupables de l'une des infractions prévues par le présent chapitre encourent également les peines complémentaires suivantes (...) $\left[l\right]$'interdiction des droits civiques, civils et de famille, (...) l'activité professionnelle ou sociale dans l'exercice ou à l'occasion de l'exercice de laquelle l'infraction a été commise, (...) $\left[et\right]$ la confiscation de la chose qui a servi ou était destinée à commettre l'infraction ou de la chose qui en est le produit.''}} governs

\begin{quote}
\textit{$\left[n\right]$atural persons guilty of one of the offenses provided for in this chapter also incur the following additional penalties: (...) $\left[t\right]$he ban on civil and family rights, (...) the professional or social activity in the exercise or on the occasion of the exercise of which the offense was committed, (...) $\left[and\right]$ the confiscation of the thing that served or was intended to commit the offense or the thing that is the product of it.}
\end{quote}
 
On the other hand, Brazil is much more advanced in this sense when compared to France, at least in theory. The brazilian legislation brings the Act 12.527/11, popularly known \textit{Access to Information Act}, promulgated in 18 November 2011 \cite{brasil2011lei} and regulated in 16 May 2012 \cite{brasil2012decreto}. It systematizes the data access in all public spheres in Brazil, and guarantees to Brazilian citizens the right to request access to public data. In the legal text\footnote{ ``\textit{É dever dos órgãos e entidades públicas promover, independentemente de requerimentos, a divulgação em local de fácil acesso, no âmbito de suas competências, de informações de interesse coletivo ou geral por eles produzidas ou custodiadas. (...) Para cumprimento do disposto no caput, os órgãos e entidades públicas deverão utilizar todos os meios e instrumentos legítimos de que dispuserem, sendo obrigatória a divulgação em sítios oficiais da rede mundial de computadores (internet). (...) Os sítios (...) deverão  (...) atender (...) aos seguintes requisitos: I - conter ferramenta de pesquisa de conteúdo que permita o acesso à informação de forma objetiva, transparente, clara e em linguagem de fácil compreensão; II - possibilitar a gravação de relatórios em diversos formatos eletrônicos, inclusive abertos e não proprietários, tais como planilhas e texto, de modo a facilitar a análise das informações; III - possibilitar o acesso automatizado por sistemas externos em formatos abertos, estruturados e legíveis por máquina; IV - divulgar em detalhes os formatos utilizados para estruturação da informação; V - garantir a autenticidade e a integridade das informações disponíveis para acesso; VI - manter atualizadas as informações disponíveis para acesso.}''} can be found the following:

\begin{quote}
\textit{It is the duty of public bodies and entities to promote, regardless of requirements, the disclosure in a place that is easily accessible, within the scope of their competences, of information of collective or general interest produced or guarded by them. (...) In order to comply with the caput, public bodies and entities shall use all legitimate means and instruments at their disposal, and disclosure on official websites of the World Wide Web (Internet) is mandatory. (...) Sites (...) shall (...) meet (...) the following requirements: I - contain content search tools that allow access to information in an objective, transparent, clear and in easy to understand language; II - enable the recording of reports in various electronic formats, including open and non-proprietary, such as spreadsheets and text, in order to facilitate the analysis of information; III - enable automated access by external systems in open, structured and machine readable formats; IV - disclose in detail the formats used for structuring the information; V - guarantee the authenticity and integrity of the information available for access; VI - keep updated the information available for access.} \cite[Article 8]{brasil2011lei}
\end{quote}

The authors, however, were unsuccessful in most data requests through official brazilian channels. In 4 December 2014 at 09:46:34, the Administrative Proceeding No. 22.031/2014 has been opened regarding the Manifestation No. 211234/2014. The manifestation mentions the Act 12.527/11 to request access to a sample database for conducting academic work. The solicitation\footnote{ \textit{a) a base de metadados relativa à base processual e b) a extração de dados públicos (brutos) processuais, com o maior histórico possível. Segue abaixo lista de informações que necessito para o estudo: nome do advogado, comarca, situação, número Themis, número CNJ, parte, classe/natureza, órgão julgador, última movimentação, acórdão, processo principal, processo de 1o grau, relator, data de distribuição, volumes, quantidades de folhas, notas de expediente, último julgamento, dados do 1o grau, depósitos judiciais e última atualização.
}} requested the following:

\begin{quote}
\textit{(a) the metadata database of the court lawsuits and (b) the extraction of court lawsuits (raw) public data with the longest possible record. Below is a list of the information I need for the study: name of lawyer, county, situation, Themis number, CNJ number, part, class/nature, judging body, latest move, judgment, main proceedings, 1st degree case, rapporteur, date of distribution, volumes, sheet quantities, file notes, last judgment, 1st degree data, court deposits and last update.}
\end{quote}

Five days later a feedback was given by TJDFT-COSIST, the First Instance Systems and Statistics Coordination\footnote{ COSIST -- \textit{Coordenadoria de Sistemas e Estatísticas da Primeira Instância}.} of the Court of Justice of the Federal District and Territories\footnote{ TJDFT -- \textit{Tribunal de Justiça do Distrito Federal e dos Territórios}.}. The document No. 22.031/2014, sent in 9 December 2014, informed about the impossibility to fulfill the request for public databases. The allegations\footnote{ \textit{No caso em questão o pedido não limita os dados desejados, e indica apenas a extração de bases de dados públicos, o que não é viável tendo em vista a quantidade de registros daquela, e a ausência de acesso desta Coordenação a tais registros para extração. Além disso, não há integração entre os sistemas informatizados da primeira e da segunda instância, de modo que a verificação de algumas informações relacionadas no presente exigirá a verificação manual dos processos. Diante do exposto, esta Coordenação informa a impossibilidade de atendimento da solicitação.}} was the following:

\begin{quote}
\textit{In this case, the request does not limit the desired data, and only indicates the extraction of public databases, which is not feasible in view of the number of records of that, and the lack of access of this Coordination to such records for extraction. In addition, there is no integration between the first and second instance computer systems, so verification of some related information here will require manual verification of the processes. Given the above, this Coordination informs the impossibility of fulfilling the request.}
\end{quote}

\begin{figure}[!h]
  \begin{center}
    \includegraphics[width=9.7cm]{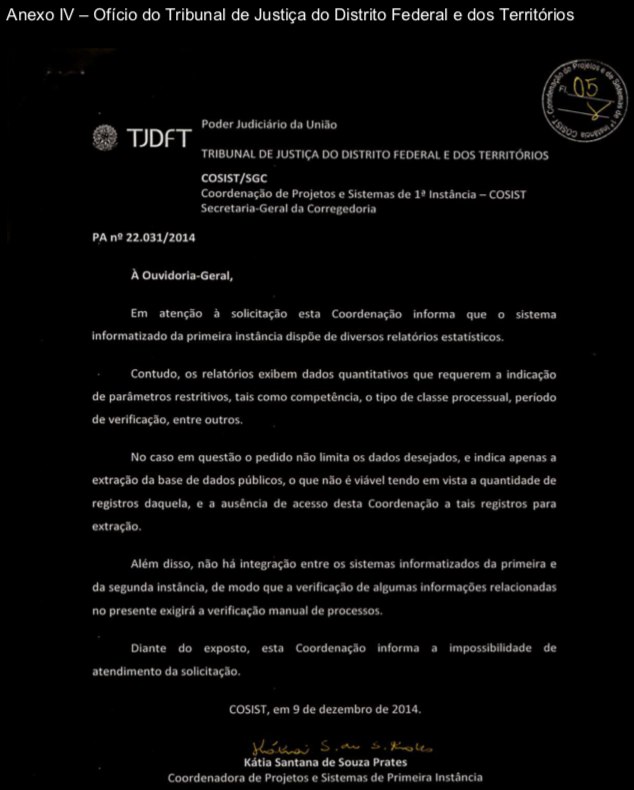}
  \caption{Administrative Proceeding No. 22.031/2014.}
  \end{center}
\end{figure}

After some attempts to gain access to public databases, the only useful feedback came from TJMG-SEPLAG-CEINFO, the Institutional Management Information Center\footnote{ CEINFO -- \textit{Centro de Informações para Gestão Institucional}.} and the Executive Secretariat of Planning and Quality in Institutional Management\footnote{ SEPLAG -- \textit{Secretaria Executiva de Planejamento e Qualidade na Gestão Institucional}.} of the Minas Gerais Court\footnote{ TJMG -- \textit{Tribunal de Justiça de Minas Gerais}, \url{http://www.tjmg.jus.br/}.}. They made available in 5 December 2014 a sample of 11 variables for 5,080,270 TJMG court lawsuits from 2000 to 2013, analyzed by \cite{oliveira2014jurimetria}. In Section \ref{sec:data} it is presented \texttt{tjmg\_year}, a clean and ready-to-use version of TJMG dataset, concerning the counting of 4,236,229 lawsuits grouped by year.

\section{Three Prisms of Jurimetrics}

\begin{quote}
\textit{Much of the activity of government, including that of lawmakers, judges, administrators and other lawyers consists of investigating and ascertaining facts.} \cite[p. 535]{loevinger1966law}
\end{quote}

The \textit{Three Prisms of Jurimetrics} is an approach proposed half decade ago by \cite{zabala2014jurimetria}. The authors intended to unify the subject, in order to give a formal connection between the three main players of the judiciary: judge, lawmaker and lawyer. As the universe of study is wide, the analysis permeates all forms of legal action and needs a didactic way to understanding the role of quantitative methods in the legal framework. The prisms are, therefore, the looks or points of view of the main agents of the judiciary. \cite{schlag2009dedifferentiation} considers the `dedifferentiation problem', in the sense that `more sophisticated theories of law lead us to a point where we are no longer able to distinguish law from culture, or society, or the market, or politics or anything of the sort'. Anyway, considering the separation of powers model -- consecrated on the Articles \nth{2} and \nth{60} of the Brazilian Federal Constitution\footnote{``\textit{Art. 2$^{\circ}$ São Poderes da União, independentes e harmônicos entre si, o Legislativo, o Executivo e o Judiciário.}''}\footnote{``\textit{Art. 60$^{\circ}$, § 4$^{\circ}$ Não será objeto de deliberação a proposta de emenda tendente a abolir: III -  a separação dos Poderes.}''} \cite{brasil988constituicao} -- and the Occam's Razor principle, we are still interested in delimiting the study of jurimetrics to the judiciary, although there may be associations of the latter with other entities.

\subsection{The Judge}

The first prism refers to the judge's view. A judge's main activity is to decide in situations under uncertainty. Considering the intersections between computing, quantitative methods and the law, it is possible to build modern calculators in order to support the magistrate's decision. \cite{degroot2004optimal} presents methods for \textit{optimal statistical decisions} considering the \textit{subjective probabilities} as numerical representations of the decision maker's beliefs and \textit{utility} as numerical representations of his/her tastes and preferences. The approach given by Morris DeGroot maximizes the expected utility taking the primitives `is more likely than' for probability and `is preferred to' for utility, in order to disentangle these two components. The opinion about a latent parameter $\theta$ is updated under the Bayesian paradigm.

Judicial expertise can be brought to the judicial process, in order to resolve doubts raised by the judge. Guidelines on the reality of precedents can also be considered, demonstrating the available information on parameters of interest. \cite{zabala2019Bayesian} `presents a new tool to support the decision concerning moral damage indemnity values of the judiciary of Rio Grande do Sul, Brazil. A Bayesian approach is given, in order to allow the assignment of the magistrate’s opinion about such indemnity amounts, based on historical values. The solution is delivered in free software using public data, in order to permit future audits'.

In addition to the reasons already presented, there is a possibility to consider the contribution of the \textit{expert witness}. The `expert testimony has become increasingly essential in a wide variety of litigated cases' \cite{berger2011admissibility}, albeit `has many virtues, and just as many negative aspects' \cite{cohen2015expert}. `The probability of winning increases with the skillful presentation of evidence', argues \cite{matson2012effective}. \cite{aitken1991use} states `the court is concerned with the probabilities (or odds) of guilt, and the expert witness is concerned with the probabilities of the evidence'. Once the opinion of an expert witness is widely accepted, it helps to resolve questions regarding the legal process involving the lack of immediate evidence or even in cases of difficult measurement such as, for instance, cases involving discrimination or prejudice.

\subsection{The Lawmaker}

The second prism refers to the lawmaker's/legislator's view. It is foreseen as one of the main activities of the legislator to purpose data-based bills, considering technical procedures and a legislative impact study. Fortunately, the Brazilian legislation is already taking this path. Know as \textit{Economic Freedom Act} \cite{brasil2019lei}, it determines that proposals for editing and amending normative acts should be accompanied by a regulatory impact analysis.\footnote{``\textit{Art. 5$^{\circ}$  As propostas de edição e de alteração de atos normativos de interesse geral de agentes econômicos ou de usuários dos serviços prestados, editadas por órgão ou entidade da administração pública federal, incluídas as autarquias e as fundações públicas, serão precedidas da realização de análise de impacto regulatório, que conterá informações e dados sobre os possíveis efeitos do ato normativo para verificar a razoabilidade do seu impacto econômico.}''}

\begin{quote}
\textit{Proposals for editing and amending normative acts in the general interest of economic agents or users of the services rendered, issued by a federal government agency or entity, including municipalities and public foundations, will be preceded by a regulatory impact analysis, which will contain information and data on the possible effects of the normative act to verify the reasonableness of its economic impact.} \cite[Art. 5]{brasil2019lei}
\end{quote}

In addition to supporting new bills and amendments to normative acts, it is required more technical forms and scientific studies to protect the population from irresponsible arbitrariness. Empirical and technical analysis helps to control the pure rhetoric and demagogy in the production of laws. This kind of approach helps to understand the areas we should prioritize efforts, focus investments and so on.

In order to helping the public administration to understand the judiciary behavior, it is possible to present realized and forecasted values from the available datasets. In Example \ref{sec:pvf} there are presented the realized values from January 2000 to December 2017, as well as the forecasting from January 2018 to July 2021. The algorithm uses only the realized values to train and test different models. Available in function \texttt{fits} from the \texttt{jurimetrics} package, it is a way to easily forecast the court proceedings volume and other univariate time series. Such analysis can be used to anticipate and improve resources allocation, such as provision of funds, agency expenses and all sort of elements related to how provisions should be made. Regarding the management of the judiciary and other public institutions, there are several examples of how the public data analysis can help in the decision making for the coming months and years. 

In Brazil, there is an explicit need for data analysis in the discussion about the Civil Procedure Code (\textit{CPC}) amendment proposals concerning court appeals suppression. The debate revolves around the following question: Does court appeals suppression significantly alter the timing of the process? To the best of our knowledge, there were no proposals considering technical analysis or somehow evaluating financial and social impacts of such suppression over time.

On the other hand, in 2015 the Brazilian Jurimetrics Association\footnote{\textit{Associação Brasileira de Jurimetria - ABJ.}} identified, in partnership with the National Council of Justice\footnote{\textit{Conselho Nacional de Justiça - CNJ.}}, critical points in the child adoption process. The findings motivated the Amendment Project no. 5850/2016 \cite{brasil2016projeto}, approved by the Brazilian congress in 22 November 2017 as the Act 13.509/2017 \cite{brasil2017lei}. This Act impacts the Statute of the Child and Adolescent (\textit{ECA}) and the Consolidation of Labor Laws (\textit{CLT}), streamlining judicial procedures related to the dismissal of family power and the adoption of children and adolescents.

The Brazilian Jurimetrics Association also made a study called `Biggest Litigants', an endeavor to identify the most frequent litigants and pointing out alternative ways to solve this class of legal problems. The study considered data from the website \href{https://www.consumidor.gov.br/}{\nolinkurl{consumidor.gov.br}} and of some Brazilian courts. The authors were honored to contribute to this study by helping to clean up the TJRS databases. The main problem was the lack of a unique identification key for each litigant. In the case, it was necessary to use REGular EXpressions (REGEX) to find patterns and group the litigants by name. According the technical report\footnote{\url{https://abj.org.br/cases/maiores-litigantes-2/}}, `only 20 companies concentrate more than 50\% of disputes', and `in the state of São Paulo 30 companies concentrate more than 70\% of the judicial processes'. The findings also indicates `the telephone companies and financial institutions consistently grouping more than 40\% of processes across all surveyed federation units. In addition, it is possible to identify that consumption ratios vary according to the location under study. Most litigation in all economic areas discuss damages for moral damages, which are often associated with improper registration in delinquent databases'.

\subsection{The Lawyer}

The third prism refers to the lawyer's view. The lawyer can use technical studies to support requirements, evaluate probabilities or even forecast values. This professional can use the quantitative analysis to measure and improve his/her performance. It is currently feasible to demonstrate on court, for instance, the risks that involve a claim if a right anticipation is not granted. Note we are not talking about the end of the legal argument, but its enrichment.

\begin{quote}
\textit{It is difficult to imagine what difference it makes to a juror, lawyer, or scientist to be told that a defendant is 218 times more likely than some man chosen at random to be the father of a child or that there is a 99.54\% probability that the defendant is the father.} \cite[p. 343]{loevinger1992standards}
\end{quote}

The Brazilian Civil Procedure Code observes the use of probability in decisions concerning urgency guardianship. Article 300 considers the danger of damage and risk to the useful result. Such concepts refer to the use of statistics and quantitative methods in general. The law is translated into probabilistic terms, indicating the legislator's intention to treat the legal issue in a technical and scientific manner.

\begin{quote}
\textit{Art. 300. The urgency guardianship will be granted when there are elements that evidence the probability of the right and the danger of damage or the risk to the useful result of the process.}\footnote{\textit{Art. 300. A tutela de urgência será concedida quando houver elementos que evidenciem a probabilidade do direito e o perigo de dano ou o risco ao resultado útil do processo.}} \cite{brasil2015lei}
\end{quote}

Another utility of quantitative approach for the benefit of law is -- as Holmes Jr., Loevinger and other researchers suggest -- the predictability of legal decision. Judgments, systematically analyzed, can provide answers about the probability of success of a claim. Once many lawsuits are filed considering these probabilities, it is important to consider this technical perspective in order to maximize the chances to bring a financial return to the lawyer or office. It happens that in the vast majority of cases the analysis of financial viability of the demand judgment is not properly analyzed, measuring risks inappropriately and often misguiding customers. Considering the analysis of the `obvious rich' court decisions, there are also several possibilities this kind of databases can bring, such as the analysis of judges profile, projection of indemnity values, estimated time of the process, among others.

In business law offices there is a need to estimate the provision of lawsuit funds. Estimated amount to be disbursed in lawsuits permits proper allocation of money for corporate judicial liabilities. Still considering law offices, one can mention the benefits that quantitative methods can bring to compliance. Jurimetrics can help in risk assessment, strategy development and internal controls, allowing for more objective and verifiable evaluations. This approach supports decisions to be taken, directing efforts towards legal security.

Another major point in legal action through the prism of advocacy -- as well other prisms -- is the possibility of alternative means for problem solving. Using state of the art technology, new platforms of agreements and arbitration tend to simplify and massify dispute resolutions. Examples of this technology are \textit{Alternative Dispute Resolution} (ADR), \textit{Online Dispute Resolution} (ODR) and \textit{smart contracts}. According to \cite[p. 23]{wang2009online}, `ODR is usually known as any of online ADR, e-ADR, iADR, virtual ADR and cyber ADR. It was technologically developed in the US and Canada, and it is still used mainly in the US'. The concept of smart contracts was coined by \cite{szabo1996smart}, and refers to a computer protocol used to intermediate trackable and irreversible transactions, and was consecrated by \cite{nakamoto2008bitcoin}.

Finally, the modern lawyer must be able to use tools that allow the practitioner to describe a set of instructions to be performed. Spreadsheets and programming languages are examples of this kind of tools. A good way to start is by understanding how spreadsheets work, and what they can do to simplify daily workflows. The most spreadsheets allow programming, usually called `macros'. The operator can record a set of instructions by using point-and-click, that is automatically converted in written language. The most widely used language in spreadsheets is BASIC and its dialects \cite{kemeny1964basic}. Besides BASIC, there are dozens of languages able to solve complex problems. A trained operator is capable to build solutions in a fast and scalable way.

\section{Applications}
 
\subsection{Tools}

The collection of tools used by a data analyst can be extensive. It will depends of the objective, scope and resources available. Usually this professional must be able to build their own tools, constantly adapting it to the dynamics of gather, analyze and presenting data. To allow a critical assessment to be carried out, however, it is necessary the considered data-related elements be available for any person.

The verification and reproducibility principles are pillars of science, and was recently discussed by \cite{pashler2012editors}, \cite{baker2016why}, \cite{munafo2017manifesto} and \cite{wasserstein2019moving}. They points out there is as crises of confidence and reproducibility, much due to a misuse of statistics, frequently the p-value misinterpretation. In order to guarantee good practices in science, projects like \textit{The Dryad Digital Repository}\footnote{ \url{https://datadryad.org}} defends open, easy-to-use, not-for-profit and community-governed data infrastructure for scholarly literature. On the same line is \textit{The Science Code Manifesto}\footnote{ \url{http://sciencecodemanifesto.org/}}, in which the signatories adopt five principles:

\begin{enumerate}
  \item \textbf{Code} All source code written specifically to process data for a published paper must be available to the reviewers and readers of the paper.
  \item \textbf{Copyright} The copyright ownership and license of any released source code must be clearly stated.
  \item \textbf{Citation} Researchers who use or adapt science source code in their research must credit the code’s creators in resulting publications
  \item \textbf{Credit} Software contributions must be included in systems of scientific assessment, credit, and recognition.
  \item \textbf{Curation} Source code must remain available, linked to related materials, for the useful lifetime of the publication.
\end{enumerate}

There is a plethora of projects based on free and open source software for those who wish to do science. Considering the data analysis and presentation, some examples are given on Table \ref{tab.free.soft}. Each software described can be extended with libraries, packages and user defined functions. For instance, recently \cite{gerum2019pylustrator} presents \texttt{Pylustrator}, an open source library tool for Python to generate the code necessary to compose publication figures from single plots. In R language \cite{r2019r} is built \texttt{ggplot2} \cite{wickham2016ggplot2}, a system for `declaratively' creating graphics based on \textit{The Grammar of Graphics} \cite{wilkinson2005grammar}. A general purpose tools for jurimetrics can be found at the \texttt{jurimetrics} package \cite{zabala2019jurimetrics}.

\begingroup
\fontsize{8pt}{8pt}\selectfont
\begin{table*}\centering
  \ra{1.3}
  \begin{tabular}{llc} \toprule
  \cmidrule{1-2} 
  Application   & URL \\ \midrule
  Cytoscape.js  & \url{http://js.cytoscape.org/}\\
  C++           & \url{http://www.cplusplus.com/}\\
  D3.js         & \url{https://d3js.org/}\\
  GNU Octave    & \url{https://www.gnu.org/software/octave/}\\
  GNU PSPP      & \url{https://www.gnu.org/software/pspp/}\\
  JASP          & \url{https://jasp-stats.org/}\\
  Julia         & \url{https://julialang.org/}\\
  \LaTeX        & \url{https://www.latex-project.org/}\\
  LibreOffice   & \url{https://www.libreoffice.org/}\\
  Markdown      & \url{https://www.markdownguide.org/}\\
  MySQL         & \url{https://www.mysql.com/}\\
  Orange        & \url{http://orange.biolab.si/}\\
  Python        & \url{https://www.python.org/}\\
  R             & \url{https://www.r-project.org/}\\
  Scilab        & \url{https://www.scilab.org/}\\
  Stan          & \url{https://mc-stan.org/}\\
  Tabula        & \url{https://tabula.technology}\\
  TensorFlow    & \url{https://www.tensorflow.org/}\\
  \bottomrule
  \end{tabular}
  \caption{A sample of free software to handle and present data}
  \label{tab.free.soft}
\end{table*}
\endgroup

R is a free software environment for statistical computing and graphics. It compiles and runs on a wide variety of \href{https://en.wikipedia.org/wiki/Unix-like}{\nolinkurl{Unix-like}} platforms and Windows. It is a \href{https://www.gnu.org/}{\nolinkurl{GNU}}\footnote{ The GNU General Public License is a type of license used for free software that grants end users (individuals, organizations or companies) the freedom to use, study, share and modify the software.} project which is similar to the S language and environment, developed at Bell Laboratories (formerly AT\&T, now Lucent Technologies) by John Chambers and colleagues. It follows a minimalist object-oriented concept, which specifies a small standard core accompanied by language extension packages. To install R the reader may access the official website project\footnote{ \url{https://cloud.r-project.org/}}. It is recommended to always keep R at its latest version as well the using of RStudio editor\footnote{ \url{https://rstudio.com/}}, an R-integrated development environment. Among many tools it enables the creation of automatic presentations and reports in various formats such as pdf, html and docx, merging languages such as R, \LaTeX, markdown, C++, Python, SQL and D3. It is available in Desktop and Server editions along with their respective previews, gathering R functionalities harmonically.

The packages used in this article can be installed and updated with the following R code. For Unix-like operating systems, it is recommended to run the following instructions on a terminal after executing the \texttt{sudo R} command followed by the admin password.

% code
\begin{tcolorbox}[colback=black, coltext=white]
\begingroup
\fontsize{9pt}{9pt}\selectfont
\begin{verbatim}
> packs <- c("tidyverse","devtools")
> install.packages(packs, dep = T)
> devtools::install_github("filipezabala/jurimetrics", force = T)
> update.packages(ask = F)
\end{verbatim}
\endgroup
\end{tcolorbox}

\subsection{Data}\label{sec:data}

\textit{Data} is a collection of tables, documents and files, usually transformed in binary format. \cite{breiman2001statistical} asserts `\textit{Statistics starts with data}', that are used `\textit{to predict and to get information about the underlying data mechanism}'. The data gathering is usually laborious, taking much of the total analysis time. Tools like MySQL, Python or R -- referenced on Table \ref{tab.free.soft} -- may help in data handling.  At this point it is considered that the reader has already understood the relevance in mastering high-level languages for the solution of applied problems. In this way it is considered that the reader performs simple operations using command line\footnote{ \url{https://en.wikipedia.org/wiki/Command-line_interface}}, the reading of technical documentation and executes other eventual tasks, although it can be time-consuming. 

Even not all applications needs substantial amount of data, it is commonplace to use a considerable structure of databases and processing. In a personal computer it is possible to create useful tools and deliver solutions for relevant problems. Some databases used by the authors are available in \texttt{jurimetrics} package.

% code
\begin{tcolorbox}[colback=black, coltext=white]
\begingroup
\fontsize{9pt}{9pt}\selectfont
\begin{verbatim}
> data("tjmg_year")
> y <- ts(data = tjmg_year$count, start = c(2000), frequency = 1)
> head(y, 10)

Time Series:
Start = 2000
End = 2009
Frequency = 1
 [1]  38403  49560  66653  81005  92012 107442 164101 213774 280847 343614

> sum(y)
[1] 4236229
\end{verbatim}
\endgroup
\end{tcolorbox}

In opposition of data collecting, some practical situations brings the necessity to decide with sample size near to (if not) zero, considering only the previous decision maker's experience and expected utility. The Bayesian approach presents tools to address solutions involving \textit{small} to \textit{big} data structures, using available data to update the opinion using the \textit{likelihood principle}. Such principle asserts the information\footnote{ According to \cite{gosh1988statistical}, in Basu's sense `\textit{information is what information does. It changes opinion. Only a Bayesian knows how to characterize his/her prior opinion on $\theta$ as a prior distribution $q(\theta)$. This prior opinion is changed, by the data $x$, to the posterior opinion $q^{*}(\theta)= q(\theta)L(\theta)/\sum q(\theta)L(\theta)$}'.} about some unknown parameter $\theta$, related to an observed variable $X$, is obtained only through the \textit{likelihood function}. The Bayesian operation calibrates the previous opinion, contained in a \textit{prior distribution} $\pi(\theta)$, with the likelihood function $L(X|\theta)$. The Bayesian then assigns his/her opinion about $\theta$ after the data to the \textit{posterior distribution} $\pi(\theta|X)$. The mathematical representation is given by \[ \pi(\theta|X) = \frac{\pi(\theta) L(X|\theta)}{\int_{\theta} \pi(\theta) L(X|\theta) d\theta} \propto \pi(\theta) L(X|\theta). \] 

\subsection{Examples}\label{sec:ex}
This section brings two examples in which the authors currently work. Code and documentation can be found at \href{https://www.github.com/filipezabala/jurimetrics}{\nolinkurl{github.com/filipezabala/jurimetrics}}.

\subsubsection{Preliminary injunction}\label{exe:pi}

\textit{Preliminary injunction}\footnote{ \textit{Antecipação de tutela}, in Portuguese.} is a legal instrument used in Brazil to provide the complainant an anticipated right, based on \textit{periculum in mora} (danger in delay) and \textit{fumus boni iuris} (likelihood of success on the merit of the case) principles \cite{grubbs2003international}. To base the arguments it is common for lawyers to use rhetoric, but it is possible to consider a quantitative approach for this purpose.

\begin{example}\label{exe:pi} (Protest restraint) Suppose a company has been incorrectly listed on a negative credit bureau. Consider also the customers make the purchase only if the company is not listed in negative bureaus. If a company has (i) an average of 1000 monthly credit bureau consultations, (ii) 10\% of the consultations converted in customers and (iii) an average ticket of \$3450.00 per client, the expected value of $X$: `loss amount per business day of negative credit bureau attribution' can be calculated by \[E(X) = \dfrac{1000 \times 0.1 \times 3450}{22} = 15681.82.\] 
\end{example}

The \texttt{jurimetrics} package brings the \texttt{exp\_loss} function, that performs the purposed calculations given the parameters \texttt{average.consult}, \texttt{prob.hire} and \texttt{average.ticket}. The function returns a list of two positions: \texttt{expected.value} and \texttt{text}. Note that \texttt{text} brings a rudiment of an automatic report, generated using the input parameters only. The reader can build their own custom functions combining pre-existing functions.

% code
\begin{tcolorbox}[colback=black, coltext=white]
\begingroup
\fontsize{9pt}{9pt}\selectfont
\begin{verbatim}
> library(jurimetrics)
> (ev <- exp_loss(average.consult = 1000, prob.hire = 0.1, average.ticket = 3450))

$expected.loss
[1] 15682

$text
[1] "The estimated loss amount per business day is $15681.82."

> paste0("The half of the expected loss is ", ev$expected.loss/2, ".")
[1] "The half of the expected loss is 7840.91."
\end{verbatim}
\endgroup
\end{tcolorbox}
\smiley

\subsubsection{Court proceedings volume forecast}\label{sec:pvf}

In the experience of the authors, a frequently asked question is about the court proceedings volume forecast. Any law operator should consider the amount of proceedings volume in the next months and years, as it impacts the entire chain of the judiciary. There is substancial purposes in literature to address the problem of forecasting values observed over time. The theme is usually referenced as \textit{time series}, and will be considered the treatment given by \cite{hyndman2018forecasting} and \cite{hyndman2018fpp2}.

Available in the \href{https://github.com/filipezabala/jurimetrics}{\nolinkurl{jurimetrics}} package, the database \texttt{tjrs\_year\_month} contains a sample of the monthly processual volume of 5,625,666 court lawsuits in a south brazilian court\footnote{ TJRS -- \textit{Tribunal de Justiça do Rio Grande do Sul}.} between January 2000 and December 2017. Based on the \textit{Access to Information Act} \cite{brasil2011lei} and its regulation \cite{brasil2012decreto}, the data were obtained from the court website\footnote{ \url{http://www.tjrs.jus.br/site/}} via web scraper written in Python language. The following code attaches the data in a tibble data frame format\footnote{ \url{https://cran.r-project.org/web/packages/tibble/vignettes/tibble.html}} with 216 rows and 2 columns, \texttt{yearMonth} and \texttt{count}.

% code
\begin{tcolorbox}[colback=black, coltext=white]
\begingroup
\fontsize{9pt}{9pt}\selectfont
\begin{verbatim}
> library(jurimetrics)
> data(tjrs_year_month) # attaching data
\end{verbatim}
\endgroup
\end{tcolorbox}

In order to print the first 10 lines, just enter the name of the attached data. Note the information about the class of the object (\texttt{tibble}) is a $216 \times 2 \times 1$ dimension tensor. The tibble object also brings information about the columns classes, respectively \texttt{<date>} and \texttt{<int(eger)>}. In the last line is displayed the number of rows and columns not shown on the screen, in this case 206 rows. The option \texttt{print(n = Inf)} prints all the lines, but this procedure is not encouraged.

% code
\begin{tcolorbox}[colback=black, coltext=white]
\begingroup
\fontsize{9pt}{9pt}\selectfont
\begin{verbatim}
> tjrs_year_month

# A tibble: 216 x 2
   yearMonth  count
   <date>     <int>
 1 2000-01-01    12
 2 2000-02-01   222
 3 2000-03-01   576
 4 2000-04-01   647
 5 2000-05-01  1098
 6 2000-06-01  2307
 7 2000-07-01   107
 8 2000-08-01  4743
 9 2000-09-01  3301
10 2000-10-01  4255
# … with 206 more rows

> # tjrs_year_month %>% print(n = Inf) # print all the 216 lines
\end{verbatim}
\endgroup
\end{tcolorbox}

As the data is time series-like, it is recommended to convert it in a \textit{time-series object} through the function \texttt{ts}. The function transforms the \texttt{data} from a \texttt{start} date, in this case January 2000 indicated by \texttt{c(2000,1)}. The \texttt{frequency} of observations must be supplied, in which 12 refers to monthly, 6 to biannual, 3 to quarterly and 1 to yearly observations. %or the number of observations per unit of time. 

At a glance it is possible to observe peaks of lawsuits judged in the months of March, April, August and September, while valleys occur in January. In this sense, we are looking for a model that capture this -- and maybe other not so obvious -- behaviors to make useful predictions. It is noteworthy, however, that all models are simplifications and idealizations of reality. The idea that complex physical, biological or social systems can be precisely described by a model is unlikely. Building idealized representations that capture important stable aspects of such systems is, however, a vital part of scientific, political and market analysis, driving decisions in situations of uncertainty. The use of modelling usually leads to helpful insights in the achievement of theoretical and practical results. \cite{box1979robustness} points out that `\textit{all models are wrong but some are useful}', suggesting
\begin{quote}
\textit{$\left[f\right]$or such a model there is no need to ask the question `Is the model true?'. If `truth' is to be the `whole truth' the answer must be `No'. The only question of interest is `Is the model illuminating and useful?'} (\cite[pp. 202-203]{box1979robustness}
\end{quote}

% code
\begin{tcolorbox}[colback=black, coltext=white]
\begingroup
\fontsize{9pt}{9pt}\selectfont
\begin{verbatim}
> (y <- ts(data = tjrs_year_month$count, start = c(2000,1), frequency = 12))

       Jan   Feb   Mar   Apr   May   Jun   Jul   Aug   Sep   Oct   Nov   Dec
2000    12   222   576   647  1098  2307   107  4743  3301  4255  5017  4819
2001   220  3153  4823  4054  5868  5084   199  6271  3958  4816  5344  5366
2002   218  3185  4793  4812  5962  6059   296  5914  5057  8068  9574 13298
2003  1293  8726 12639 14259 14241 16940   953 18850 17196 20102 18526 20440
2004   754 11137 24722 20646 20489 25490  3686 23938 23300 21315 21965 25790
2005   760 11998 26251 22408 21814 29089 18033 26014 23623 22118 25293 25723
2006 15794 12219 27052 25350 28483 28662 23002 34301 27000 28537 32323 31973
2007  7171 21820 34138 29551 34852 30634 26801 36776 30171 34086 32273 29654
2008 12780 20196 32816 34032 33247 35545 34177 45412 38270 45866 39761 33548
2009 18204 17460 43022 43083 35652 41244 38571 42821 43666 40700 39835 32906
2010 22982 20864 43883 38889 44821 37571 39317 40314 38927 38610 39392 34991
2011 20010 27162 37795 36449 38173 42342 35974 44597 40687 37855 37043 34255
2012 17173 24373 42825 37704 45831 40870 32243 45656 34618 43379 37585 32612
2013 16343 21962 38792 39893 38117 39811 31737 41841 36519 43293 38288 31896
2014 16136 22415 33838 36927 36900 26766 35550 40608 38130 42740 37506 37183
2015 17096 29100 41133 39881 33891 36962 36145 38498 38599 21560 44918 31096
2016 13455 23148 35486 29451 32326 43092 37561 42904 38194 38019 37958 31796
2017  8626 27212 44352 32129 36362 33478 33534 41154 32806 37405 36977 28945

> sum(y)
[1] 5625666
\end{verbatim}
\endgroup
\end{tcolorbox}

The tidy data can then be applied in the function \texttt{fits} in order to adjust the best model from classes \href{https://otexts.com/fpp2/arima.html}{\nolinkurl{ARIMA}}, \href{https://otexts.com/fpp2/ets.html}{\nolinkurl{ETS}}, \href{https://otexts.com/fpp2/complexseasonality.html}{\nolinkurl{TBATS}} and \href{https://otexts.com/fpp2/nnetar.html}{\nolinkurl{NNETAR}}. The strategy is to fit/train models with a percentage corresponding to the first points of the time series and comparing the remaining last points with the prediction made by each considered model. The comparison is made using the \textit{mean squared error} (MSE) of prediction, given by \[ MSE = \frac{1}{n} \sum_{i=1}^n (y_i - \hat{y}_i)^2\] where $n$ is the number of remaining/test points of the time series, $y_i$ is the $i^{th}$ tested value and $\hat{y}_i$ the $i^{th}$ predicted value by a given model. 

The parameter \texttt{train} is set by default to 0.8 = 80\%, remaining around 20\% of the points to test. The model that leads to prediction with the smallest MSE is declared the best model. The function returns a list with four positions: \texttt{\$fcast}, \texttt{\$mse.pred}, \texttt{\$best.model} and \texttt{\$runtime}. The \texttt{\$fcast} contains the predicted time series using the model that minimizes the MSE. \texttt{\$mse.pred} shows the MSE for each considered model, and \texttt{\$best.model} specifies the best model, in this case from the class neural network autoregressive. As suggested in Example \ref{exe:pi}, it is possible to arrange the functions to compose automatic reports.

%%%% Reader code
\begin{tcolorbox}[colback=black, coltext=white]
\begingroup
  \fontsize{9pt}{9pt}\selectfont
\begin{verbatim}
> (fit <- fits(y, show.main.graph = F))

$fcast
       Jan   Feb   Mar   Apr   May   Jun   Jul   Aug   Sep   Oct   Nov   Dec
2018 14148 28343 39777 34743 37086 35660 35452 38980 35254 37454 37383 32396
2019 17767 29523 38800 36150 37260 36734 36541 38143 36520 37473 37510 34842
2020 21250 30964 38464 36825 37340 37215 37088 37798 37105 37498 37545 36256
2021 25157 33409 38250 37172 37397 37418 37344             

$mse.pred
  mse.pred.aa mse.pred.ets mse.pred.tb mse.pred.nn
1    1.13e+08     68219968    68947440    31652653

$best.model
[1] "nnetar"

$runtime
Time difference of 3.019 secs
\end{verbatim}
\endgroup
\end{tcolorbox}

% paste0('The best model is ', fit$best.model, ', with MSE ', floor(min(fit$mse.pred)), '.')
The parameter \texttt{steps} is set by default to \texttt{NULL}, in this case projecting the number of points corresponding to the complementary percentage defined in \texttt{train}. The user may set \texttt{steps} to a value greater or equal than \texttt{train*length(x)}, never smaller. The \texttt{max.points} limits the maximum number of points to be used, set to 500 by default. If the parameter \texttt{show.main.graph} is set to \texttt{TRUE}, the main plot is presented, and \texttt{show.sec.graph = TRUE} prints the intermediate plots. If \texttt{show.value = TRUE}, the system prints the textual results of the function. The \texttt{PI = TRUE} prints the prediction intervals used in \texttt{nnar} models, but may take long time processing. Finally, \texttt{theme\_doj = TRUE} will use the Decades Of Jurimetrics theme, based on \texttt{ggplot2::theme}.

\begin{figure}[ht]
\begin{center}
\includegraphics[width=0.6\textwidth]{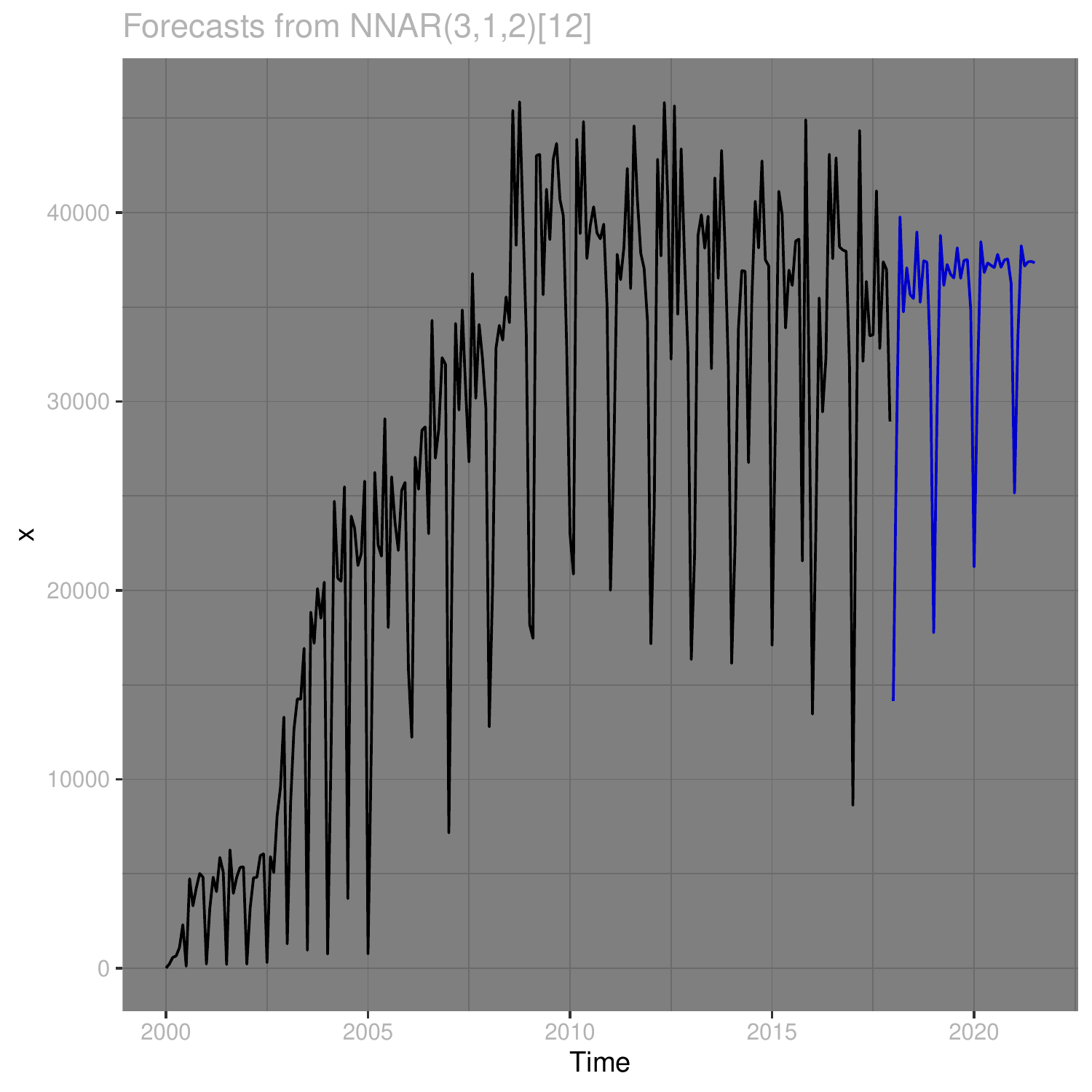}
\caption{Monthly proceedings volume forecast in TJRS Court.}
\label{fig:pvf}
\end{center}
\end{figure}

As shown in Figure \ref{fig:pvf}, the black line indicates the observed points, meanwhile the blue line indicates the predicted values. The $NNAR(3,1,2)\left[ 12 \right]$ model has inputs $y_{t-1}$, $y_{t-2}$, $y_{t-3}$ and $y_{t-12}$, and two neurons in the hidden layer\footnote{ According to \cite{hyndman2018forecasting}, a $NNAR(p,P,k)\left[ m \right]$ model has inputs  $y_{t-1}$, $y_{t-2}$, $\ldots$, $y_{t-p}$, $y_{t-m}$, $y_{t-2m}$, $\ldots$, $y_{t-Pm}$ and $k=(p+P+1)/2$ neurons in the hidden layer. A $NNAR(p,P,0) \left[ m \right]$ model is equivalent to an $ARIMA(p,0,0)(P,0,0)\left[ m \right]$ model but without the restrictions on the parameters that ensure stationarity.}. This means to make the predictions the model looks 1, 2, 3 and 12 months in the historical series, capturing monthly, bimonthly, quarterly and yearly behaviors.

%%%% Plot

\newpage
\section{The Posterior Step}\label{sec:tps}

\begin{quote}
\textit{There was a man in our town \\
and he was wondrous wise: \\
he jumped into a BRAMBLE BUSH \\
and scratched out both his eyes— \\
and when he saw that he was blind, \\
with all his might and main \\
he jumped into another one \\
and scratched them in again.} 
\end{quote}
% Llewellyn (1930) - The Bramble Bush - On Our Law and Its Study
% \cite[p. 6]{llewellyn1930bramble}

\subsection{Ignorance}

In the morning a children ask: \textit{is the man wondrous wise?}
\vspace{.5cm}

Two states may be considered: 
\[
  \begin{array}{l}
      W: \text{the man \textit{is} wondrous wise} \\ 
      \overline{W}: \text{the man \textit{is not} wondrous wise} \\ 
  \end{array} \]
      
% the man \textit{is} wondrous wise ($W$) or the man \textit{is not} wondrous wise ($\overline{W}$). 
As Dennis Lindley conveniently points out on the foreword of \cite[p. ix]{finetti1974theory}, `we shall all be Bayesian by 2020'. In advance, \cite{ferrie2019Bayesian} purposes a `babyesian' approach. So, trying to answer the children's question under this point of view, initially equal weights are assigned to the two states, let's say `fifty-fifty'. This is called \textit{Bayes-Laplace prior} and can be represented by \[ P(W) = \frac{50}{100} \hspace{2cm} P(\overline{W}) = \frac{50}{100} \] Under this perspective it is possible to answer `I'm 50\% sure that the man is wondrous wise'.

% Considering we don't know the man and don't have preference,   Let's learn about $W$ and $\bar{W}$.

\subsection{Learning}

In the afternoon, a children ask: \textit{is the man wondrous wise?}
% Unconvinced, a children ask again: \textit{is the man wondrous wise?}
\vspace{.5cm}

Even belonging to a privileged environment, it is hard to believe that 50\% of people are wondrous wise. Considering such an environment, the guess may turn around, let's say... 20-80. It can be represented by \[ P(W) = \frac{20}{100} \hspace{2cm} P(\overline{W}) = \frac{80}{100} \] Under this perspective it is possible to answer `I'm 20\% sure that the man is wondrous wise'.

Reading the poem again, it can be noted the man deliberately jumped into a bramble bush and scratched out both his eyes, going blind. Considering this information, the guess is updated to 10-90. It can be represented by \[ P(W) = \frac{10}{100} \hspace{2cm} P(\overline{W}) = \frac{90}{100} \] Under this perspective it is possible to answer `I'm 10\% sure that the man is wondrous wise'.

Continuing to read the poem, it can be noted the man jumped deliberately into a bramble bush \textit{again}. Under this perspective, the guess is updated to 5-95. It can be represented by \[ P(W) = \frac{5}{100} \hspace{2cm} P(\overline{W}) = \frac{95}{100} \] Under this perspective it is possible to answer `I'm 5\% sure that the man is wondrous wise'.

Googling it is found a Wikipedia article\footnote{\url{https://en.wikipedia.org/wiki/There_Was_a_Man_in_Our_Town}} called `There Was a Man in Our Town', indicating the poem is originally an English nursery rhyme\footnote{\url{https://en.wikipedia.org/wiki/Nursery_rhyme}}. The article reports `(i)t is believed to be based on the Greek myth of Bellerophon'\footnote{\url{https://en.wikipedia.org/wiki/Bellerophon}}. Assuming this, the guess is updated to 40-60. It can be represented by \[ P(W) = \frac{40}{100} \hspace{2cm} P(\overline{W}) = \frac{60}{100} \] Under this perspective it is possible to answer `I'm 40\% sure that the man is wondrous wise'. 

Continuing the reading, it is found Bellerophon bravely captured Pegasus and slaved Chimera. The guess is updated to 70-30. It can be represented by \[ P(W) = \frac{70}{100} \hspace{2cm} P(\overline{W}) = \frac{30}{100} \] Under this perspective it is possible to answer `I'm 70\% sure that the man is wondrous wise'. 

Continuing to read the article, it is informed Bellerophon was arrogant, angered Zeus, lived out his life in misery and... had fallen into a thorn bush causing him to become blind. The guess is then updated to 1-99. It can be represented by \[ P(W) = \frac{1}{100} \hspace{2cm} P(\overline{W}) = \frac{99}{100} \] Under this perspective it is possible to answer `I'm 1\% sure that the man is wondrous wise'. 

% Learning is the capability of listen, process and decide better.
% achieve a more accurate decision process about an issue. Let's try human learning. 

% Is the man who jumps into a bramble bush -- twice -- wondrous wise?

% In Brazil we have some denominations for somebody who jumps into a bramble bush -- twice --, and this denominations can't be translated as `wise'.

\subsection{Knowledge}

At night, a children ask: \textit{is the man wondrous wise?}
\vspace{.5cm}

Before you answer, you unintentionally bump into in a book called \textit{The Bramble Bush}. The subtitle is \textit{Some Lectures on Our Law and Its Study}. The author is \textit{Karl N. Llewellyn}, referenced as \cite{llewellyn1930bramble}. What is your posterior step?

\bibliography{doj}  % associado ao arquivo: 'doj.bib'
\bibliographystyle{plainnat}

\end{document}